\documentclass[iop,apjl,numberedappendix,deluxetable]{emulateapj}
\usepackage[latin1]{inputenc}
\usepackage{amssymb}
\usepackage{amsmath}
\usepackage{graphicx}
\usepackage{xspace}
\usepackage{natbib}
\usepackage{url}
\usepackage{paralist}
\usepackage{epsfig}

\def\cm{{\rm\thinspace cm}}

\def\erg{{\rm\thinspace erg}}
\def\eV{{\rm\thinspace eV}}

\def\keV{{\rm\thinspace keV}}
\def\km{{\rm\thinspace km}}

\def\Mpc{{\rm\thinspace Mpc}}
\def\Msun{\hbox{$\rm\thinspace M_{\odot}$}}

\def\s{{\rm\thinspace s}}

\def\yr{{\rm\thinspace yr}}

\def\ergcmps{\hbox{$\erg\cm\s^{-1}\,$}}

\def\ergpcmsqps{\hbox{$\erg\cm^{-2}\s^{-1}\,$}}

\def\ergps{\hbox{$\erg\s^{-1}\,$}}

\def\kmps{\hbox{$\km\s^{-1}\,$}}

\def\Msunpyr{\hbox{$\Msun\yr^{-1}\,$}}

\def\pcmsq{\hbox{$\cm^{-2}\,$}}

\def\h18{\hbox{H1821$+$643\,}}

\slugcomment{ }

\shorttitle{NuSTAR observations of Cygnus-A}
\shortauthors{C.~S.~Reynolds et al.}

\begin{document}

\title{{\it NuSTAR} observations of the powerful radio-galaxy Cygnus~A}

\author{Christopher~S.~Reynolds\altaffilmark{1,2},
Anne~M.~Lohfink\altaffilmark{3},
Patrick~M.~Ogle\altaffilmark{4},
Fiona~A.~Harrison\altaffilmark{5},
Kristin~K.~Madsen \altaffilmark{5},
Andrew~C.~Fabian\altaffilmark{3},
Daniel~R.~Wik\altaffilmark{6},
Grzegorz~Madejski\altaffilmark{7},
David~R.~Ballantyne\altaffilmark{8},
Steven~E.~Boggs\altaffilmark{9},
Finn~E.~Christensen\altaffilmark{10},
William~W.~Craig\altaffilmark{9,11},
Felix~Fuerst\altaffilmark{5},
Charles~J.~Hailey\altaffilmark{12},
Lauranne~Lanz\altaffilmark{4},
Jon~M.~Miller\altaffilmark{13},
Cristian~Saez\altaffilmark{1},
Daniel~Stern\altaffilmark{14},
Dominic~J.~Walton\altaffilmark{5},
and William~Zhang\altaffilmark{6}}
\altaffiltext{1}{Department of Astronomy, University of Maryland, College Park, MD 20742-2421, USA; chris@astro.umd.edu}
\altaffiltext{2}{Joint Space-Science Institute (JSI), College Park, MD 20742-2421, USA}
\altaffiltext{3}{Institute of Astronomy, Madingley Road, Cambridge, CB3 OHA, UK}
\altaffiltext{4}{Infrared Processing and Analysis Center, California Institute of Technology, MC100-22, Pasadena, CA 91125}
\altaffiltext{5}{Cahill Center for Astronomy and Astrophysics, , California Institute of Technology, Pasadena, CA 91125}
\altaffiltext{6}{Astrophysics Science Division, NASA Goddard Space Flight Center, Greenbelt, MD 20771, USA}
\altaffiltext{7}{Kavli Institute for Particle Astrophysics and Cosmology, SLAC National Accelerator Laboratory, Menlo Park, CA 94025, USA}
\altaffiltext{8}{Center for Relativistic Astrophysics, School of Physics, Georgia Institute of Technology, 837 State Street, Atlanta, GA 30332-0430, USA}
\altaffiltext{9}{Space Sciences Laboratory, University of California, Berkeley, CA 94720, USA}
\altaffiltext{10}{DTU Space, National Space Institute, Technical University of Denmark, Elektrovej 327, DK-2800 Lyngby, Denmark}
\altaffiltext{11}{Lawrence Livermore National Laboratory, Livermore, CA 94550, USA}
\altaffiltext{12}{Columbia Astrophysics Laboratory, Columbia University, New York, NY 10027, USA}
\altaffiltext{13}{Department of Astronomy, University of Michigan, 500 Church Street, Ann Arbor, MI 48109-1042, USA}
\altaffiltext{14}{Jet Propulsion Laboratory, California Institute of Technology, Pasadena, CA 91109, USA}

\begin{abstract}
\noindent We present {\it NuSTAR} observations of the powerful radio galaxy
Cygnus~A, focusing on the central absorbed active galactic
nucleus (AGN).  Cygnus~A is embedded in a cool-core galaxy cluster,
and hence we also examine archival {\it XMM-Newton} data to
facilitate the decomposition of the spectrum into the AGN and
intracluster medium (ICM) components. {\it NuSTAR} gives a
source-dominated spectrum of the AGN out to $>70$\,keV.  In gross
terms, the {\it NuSTAR} spectrum of the AGN has the form of a
power-law ($\Gamma\sim 1.6-1.7$) absorbed by a neutral column density
of $N_H\sim 1.6\times 10^{23}\pcmsq$.  However, we also detect
curvature in the hard ($>10\keV$) spectrum resulting from reflection
by Compton-thick matter out of our line-of-sight to the X-ray source.
Compton reflection, possibly from the outer accretion disk or
obscuring torus, is required even permitting a high-energy cutoff
in the continuum source; the limit on the cutoff energy is $E_{\rm
cut}>111\keV$ (90\% confidence). Interestingly, the absorbed power-law
plus reflection model leaves residuals suggesting
the absorption/emission from a fast ($15,000-26,000\kmps$), high
column-density ($N_W>3\times 10^{23}\pcmsq$), highly ionized ($\xi\sim
2,500\ergcmps$) wind.  A second, even faster ionized wind component is
also suggested by these data.  We show that the ionized wind likely
carries a significant mass and momentum flux, and may carry sufficient
kinetic energy to exercise feedback on the host galaxy.  If confirmed,
the simultaneous presence of a strong wind and powerful jets in
Cygnus~A demonstrates that feedback from radio-jets and
sub-relativistic winds are not mutually exclusive phases of AGN
activity but can occur simultaneously.
\end{abstract}



\section{Introduction}\label{intro}

Ever since its discovery in the early days of radio astronomy, the powerful radio galaxy Cygnus~A \citep{hargrave:74a} has been an important proving ground for our models of active galactic nuclei (AGN) and the relativistic jets that they produce.  It is extremely radio luminous, with a 178\,MHz luminosity more than an order of magnitude larger than any other 3C source in the local ($z<0.1$) universe \citep{carilli:96a} --- indeed, one needs to reach out to the $z\sim 1$ universe before finding many other sources of comparable radio luminosity.  Much of the radio emission originates from two edge-brightened radio lobes that are fed by relativistic and highly collimated back-to-back jets produced by the AGN, making Cygnus~A the archetypal example of a Fanaroff-Riley Type-II \citep[FRII;][]{fanaroff:74a} classical double radio galaxy.

Studies of the central engine of Cygnus~A are hampered by heavy obscuration.  The discovery of a broad MgII line in the ultraviolet spectrum \citep{antonucci:94a} and broad H$\alpha$ in the polarized optical spectrum \citep{ogle:97a} makes it clear that the central engine of Cygnus~A is a broad-line quasar, as previously suggested by \cite{djorgovski:91a} on the basis of infrared (IR) imaging.   Interestingly, Cygnus~A appears to be only a modestly powerful quasar despite powering an ultra-luminous radio-source \citep{barthel:96a}.   A recent demonstration of this was given by \cite{privon:12a} who model the radio-to-IR spectral energy distribution (SED), concluding that the IR luminosity is dominated by the AGN with a bolometric luminosity of $L_{\rm bol}\approx 4\times 10^{45}\ergps$, typical of a low-power quasar.  For a black hole mass of $M=(2.5\pm 0.7)\times 10^9\Msun$, determined via {\it Hubble Space Telescope} spectroscopy of a nuclear gas disk \citep{tadhunter:03a}, this bolometric luminosity implies a modest Eddington ratio of $L/L_{\rm Edd}\approx 0.01$.  Substantially more power is believed to emerge in a kinetic form associated with the relativistic jets; using the dynamics of the cocoon, \cite{ito:08a} estimate a total jet power in the range $L_j\sim (0.7-4)\times 10^{46}\ergps$.   An additional reason for the anomalous radio-loudness of Cygnus~A is its unusual environment --- it is hosted by the cD galaxy of a cooling core cluster of galaxies \citep{arnaud:84a,reynolds:96a,barthel:96a,smith:02a}.  The high pressure core of the intracluster medium (ICM) provides a working surface for the jets and confines the radio lobes, thereby greatly increasing the synchrotron emissivity of the shocked jet plasma.  

Hard X-ray spectroscopy is a powerful way to probe the central engines of deeply buried AGN such as Cygnus~A, although the presence of the X-ray luminous ICM complicates such studies.  While the X-ray emission from the Cygnus~A cluster was detected by the {\it Einstein} Observatory \citep{arnaud:84a}, the harder spectral response of the medium-energy (ME) telescope on {\it EXOSAT} was required to first detect the additional hard X-ray emission from the absorbed nucleus \citep{arnaud:87a}.  {\it Ginga} permitted a robust detection of the AGN continuum out to 20\,keV \citep{ueno:94a}; the continuum was well described by a power-law with photon index $\Gamma=2.0\pm 0.2$ and 2--10\,keV intrinsic luminosity of $1\times 10^{45}\ergps$ absorbed by a neutral column density of $N_H=(3.8\pm 0.7)\times 10^{23}\pcmsq$.  This basic picture, albeit with a somewhat flatter photon index ($\Gamma\approx 1.5$), was confirmed by \cite{young:02a} who used a combination of {\it Chandra} and the  {\it Rossi X-ray Timing Explorer (RXTE)} to map the AGN emission out to 100\,keV.  By using its superior spatial resolution to isolate the nuclear emission, {\it Chandra} found X-ray reprocessing of the AGN continuum in the form of a weak 6.4\,keV fluorescent line of cold iron.   However, the non-imaging nature of the instruments on {\it RXTE} meant that the hard X-ray spectra were strongly background dominated.   

In this paper, we present observations of Cygnus~A by the {\it Nuclear Spectroscopic Telescopic Array} \citep[{\it NuSTAR}; ][]{harrison:13a}.  The imaging capability of this focusing X-ray observatory, with a half-power diameter of just 60\,arcsec \citep{madsen:15a}, allows us to produce a high signal-to-noise (S/N), source-dominated spectrum of Cygnus~A out to almost 80\,keV.   The unprecedented quality of this spectrum allows us  to search for the presence of Compton reflection signatures, constrain any high-energy cutoff of the continuum, and search for highly ionized outflows  for the first time in this keystone object.  

An important complication, however, is the contribution of the ICM emission in the softest bands of {\it NuSTAR} and the all important iron K-shell band.  Thus, our {\it NuSTAR} analysis must be informed by additional soft X-ray (0.5--10\,keV) imaging-spectroscopy that allows us to construct a spectral model for the ICM.  At first glance, the moderately deep {\it Chandra} observations (totaling 200\,ks) of Cygnus-A \citep{smith:02a} would appear to be the datasets of choice for this exercise.  However, the AGN itself suffers extreme photon pile-up during these observations (compromising the integrity of any global ICM+AGN spectrum), and the hard X-ray wings of the {\it Chandra} PSF noticeably contaminate the inner 10\,arcsec of the ICM with AGN emission (compromising the ability to spatially isolate the ICM emission).  Thus, in order to obtain the best constraint on the ICM emission contaminating our {\it NuSTAR} view of the nucleus, we turn to archival {\it XMM-Newton} observations.  

This paper is organized as follows.   The {\it NuSTAR} and {\it XMM-Newton } observations and the basic data reduction steps are described in Section~2.  Section~3 gives a preliminary discussion of the {\it NuSTAR} image of Cygnus~A, although we defer a detailed imaging analysis, including the search for extended hard X-ray emission from non-thermal particles, to a future publication.  Section~4 presents our analysis of the {\it XMM-Newton} data.  With the {\it XMM-Newton} ICM model in hand, we describe the analysis of the {\it NuSTAR} spectrum in Section~5.   Section~6 discusses the astrophysical implications of our results.  Throughout this work, we assume the standard {\it Planck} cosmology \citep{ade:13a}.  At a redshift of $z=0.056$, this places Cygnus~A at a luminosity distance of 237\Mpc, with a linear-angular conversion of 1.0\,kpc per arcsec.

\section{Observations and data reduction}

\subsection{NuSTAR}

{\it NuSTAR} observed Cygnus~A on 17/18-Feb-2013 and 1-Mar-2013 with a total observation length of 80\,ks and 40\,ks respectively.   The data from both focal plane modules (FPMA and FPMB) and both observations were reprocessed and cleaned using the most recent version of the {\it NuSTAR} pipeline (within HEASOFTv16.6) and calibration files (CALDB version 20140715), resulting in 43.6\,ks and 20.7\,ks of good on-source exposure for the first and second observation, respectively.  

\begin{figure}
\centerline{
\psfig{figure=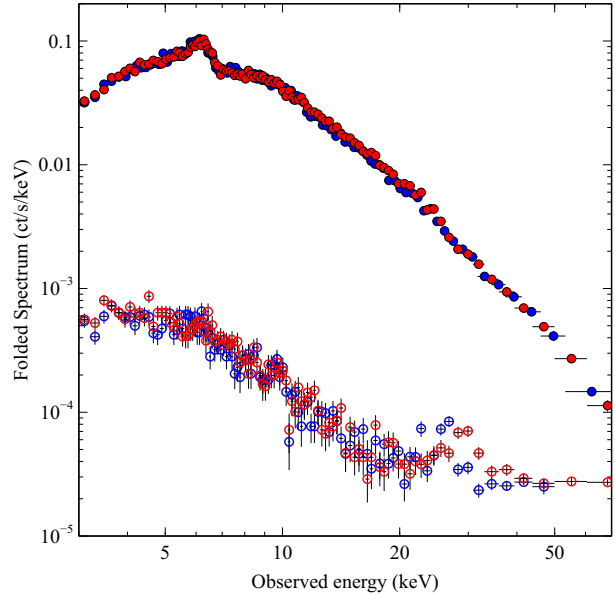,width=0.45\textwidth}
}
\caption{Folded background-subtracted spectra from the FPMA (red filled circles) and FPMB (blue filled circles), along with the background spectra (FPMA background shown as red open circles and FPMB as blue open circles).}
\label{fig:data_and_background}
\vspace{1cm}
\end{figure}

\begin{figure*}[t]
\psfig{figure=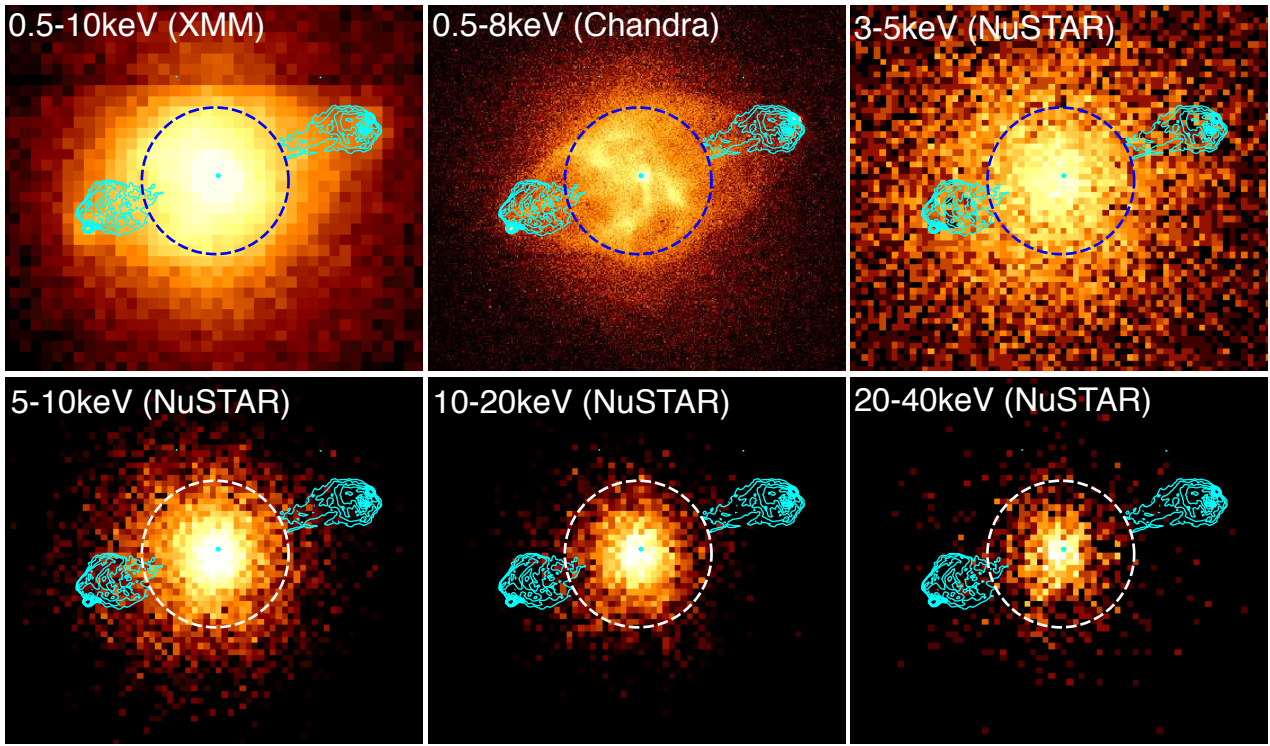,angle=90,width=0.98\textwidth}
\vspace{-1.5cm}
\caption{X-ray images of the Cygnus~A field overlaid with contours of the \cite{perley:84a} 6\,cm radio map (cyan) and our {\it XMM-Newton}/{\it NuSTAR} spectral extraction region (dashed circle of 30\arcsec radius).  Each frame has been scaled with a logarithmic color map spanning a dynamic range of 100.  }
\label{fig:images}
\end{figure*}

Although we briefly discuss the imaging data in the next section, this paper focuses on the AGN at the heart of Cygnus~A.  We extracted spectra and light curves from a 30\,arcsec radius circular region centered on the nucleus (dashed circle in Fig.~\ref{fig:images}).  Our chosen extraction region is somewhat smaller than usual in order to minimize the contribution of the ICM.   Background spectra and light curves were extracted from two circular regions that are free of any obvious point sources and flank the nucleus by 4\,arcmin.   To remain within the well-calibrated regime, this paper considers the 3--70\,keV {\it NuSTAR} spectrum and we rebin the data to a minimum of 20 photons per energy channel in order to facilitate the use of $\chi^2$-statistics.  The (folded) background-subtracted source spectra for the two FPM are shown in Fig.~\ref{fig:data_and_background} along with the corresponding background spectra.  We see that the spectra are source-dominated across the whole 3--70\,keV --- the background contributes only 5\% at 7\,keV increasing to 25\% at 70\,keV. 

Examination of the background-subtracted light curves reveals no evidence for time variability within an observation; this has been tested with lightcurves employing 500\,s, 1000\,s, and 2000\,s bins.  This result stands, even when we restrict our attention to energies above 10\,keV (thereby largely removing the contribution of the constant ICM component).   Additionally, between the two observations, the average background-subtracted 3--70\,keV count rate of each FPM is found to be constant within errors.  Hence, we combine the data from the two observations to produce a single spectrum (for each FPM) with a total exposure time of 64.3\,ks.     There is a 1.7\% offset in the count rate between the two FPMs (with count rates of $0.668\pm 0.003$\,cps for FPMA and $0.656\pm 0.003$\,cps for the FPMB) that we attribute to residual errors in the flux calibration of the two instruments.  In all of the spectral fitting performed in this paper, we fit for the cross-normalization of the two FPMs thereby removing this cross-calibration error.

\subsection{XMM-Newton}

{\it XMM-Newton} observed Cygnus~A on 14-Oct-2005 (22.5\,ks) and 16-Oct-2005 (18.8\,ks).  Here we consider just the imaging spectroscopy data from the EPIC-pn detector.   These data were extracted from the HEASARC archives and reprocessed using the {\tt epchain} tool within SASv13.5.0.  After applying standard filtering criteria (described in the {\it XMM-Newton} ABC guide), the first and second observations yielded 18.3\,ks and 15.0\,ks of good on-source data, respectively.

Our motivation for examining the EPIC-pn data is to give context to our {\it NuSTAR} study.  Hence, for each of the two observations, we extract spectra from the {\it NuSTAR} extraction radius, a 30\,arcsec radius circular region centered on the nucleus.  Background spectra were extracted from a region approximately 4\,arcmin from the nucleus that is free from any obvious point sources.  The source and background spectra were appropriately scaled (using the {\tt backscale} tool), and response matrices and effective area curves generated (using the {\tt rmfgen} and {\tt arfgen} tools).  Finally, given the proximity in time of the two observations and the lack of any detected variability between these two observations, the two spectra were combined into a single spectrum with an exposure of 33.3\,ks.  To remain within the well-calibrated regime, this paper considers the 0.5--10\,keV EPIC-pn spectrum and we rebin the data to a minimum of 20 photons per energy channel in order to facilitate the use of $\chi^2$-statistics.  

\section{Images}

A detailed study of the hard X-ray imaging data for Cygnus~A, including constraints on non-thermal X-ray emission from the Cygnus~A cluster, is deferred to a future paper.  Here, we present a preliminary discussion of the imaging data to the extent required to set up the analysis of the nuclear spectrum.   Figure~\ref{fig:images} shows the 0.5--10\,keV {\it XMM-Newton}/EPIC-pn image of the Cygnus~A along with {\it NuSTAR} images made in four bands (3--5\,keV, 5--10\,keV, 10--20\,keV and 20--40\,keV).   For reference, we also show the 0.5--5\,keV {\it Chandra}/ACIS image\footnote{This image is formed by merging the level-2 events files from ObsIDs 5830, 5831, 6225, 6226, 6228, 6229, 6250, and 6252.}.  All frames are overlaid with contours of the classic 6cm {\it Very Large Array} (VLA) map of \cite{perley:84a} as well as the {\it NuSTAR} extraction radius (dashed circle).  Each frame has been scaled with a logarithmic color map spanning a dynamic range of 100.

The {\it Chandra} image clearly shows the prolate ellipsoidal cocoon blown by the radio galaxy activity \citep{wilson:06a}.  The emission visible beyond this cocoon is the ambient ICM.   As previously noted by \cite{smith:02a}, the cocoon appears to be wrapped with X-ray bright filaments, the origin of which is unclear.  We see that the {\it NuSTAR} extraction radius fits within the radio lobes and encompasses the inner portions of the jet-blown cocoon.   In the {\it XMM-Newton}/EPIC-pn image, the cocoon structure is less distinct but the X-ray emission from the radio hot-spots is easily visible.   In {\it NuSTAR}, as expected, the extended cluster emission is also prominent in the 3--5\,keV image.  However, as can be see from the bottom panels of Fig.~\ref{fig:images}, the nuclear point source becomes increasingly dominant as one considers harder and harder bands\footnote{We note that there is only a weak energy dependence to the point spread function of {\it NuSTAR} \citep{madsen:15a}, with a half-power diameter increasing by approximately 10\% at the softest energies ($<4.5\keV$).}.   As we will see, this is completely in line with our spectral analysis.

\section{{\it XMM-Newton} spectral constraints on the AGN and inner ICM}
\label{sec:xmmanal}

\begin{figure*}[t]
\hbox{
\psfig{figure=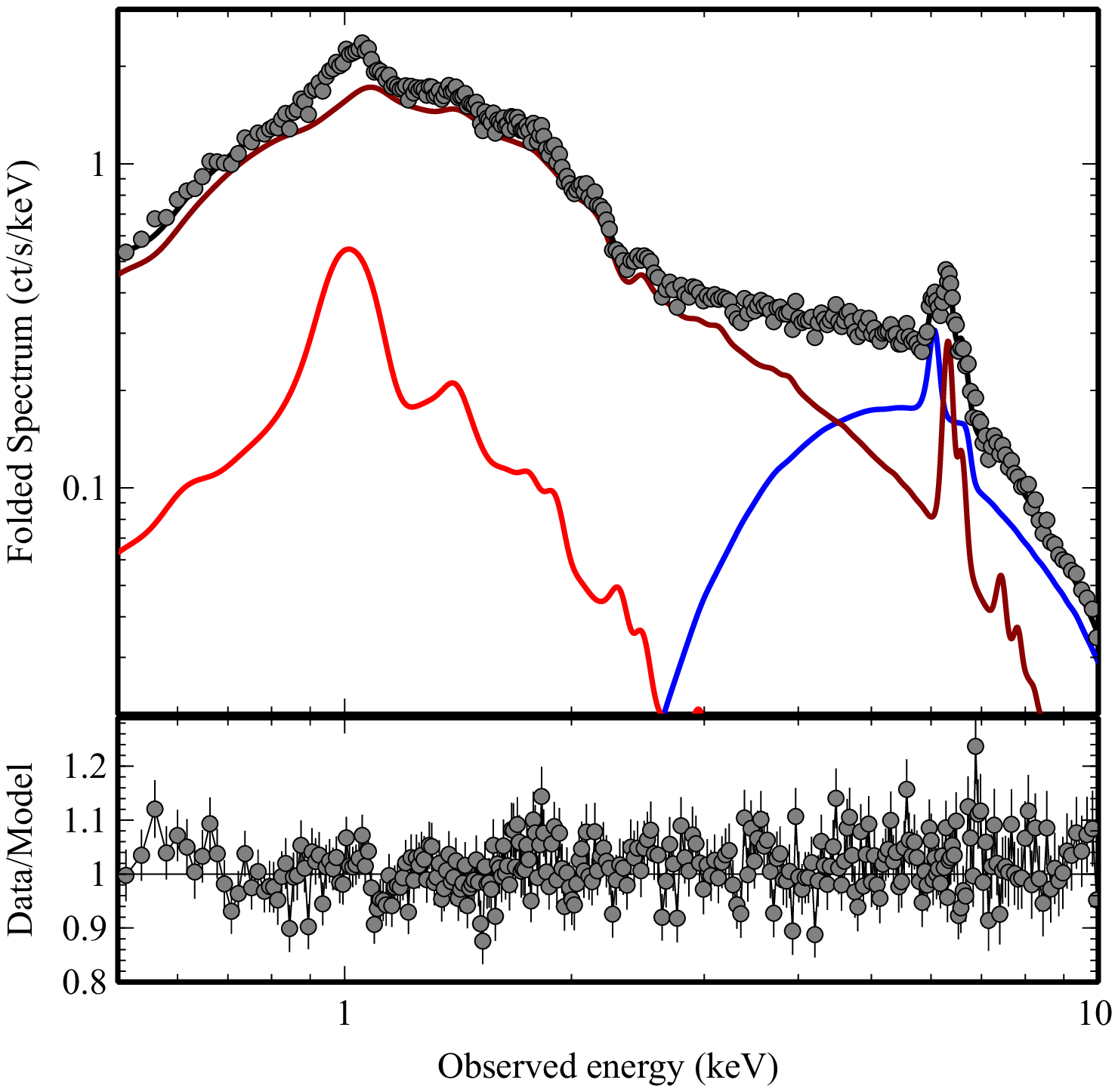,width=0.45\textwidth}
\hspace{0.5cm}
\psfig{figure=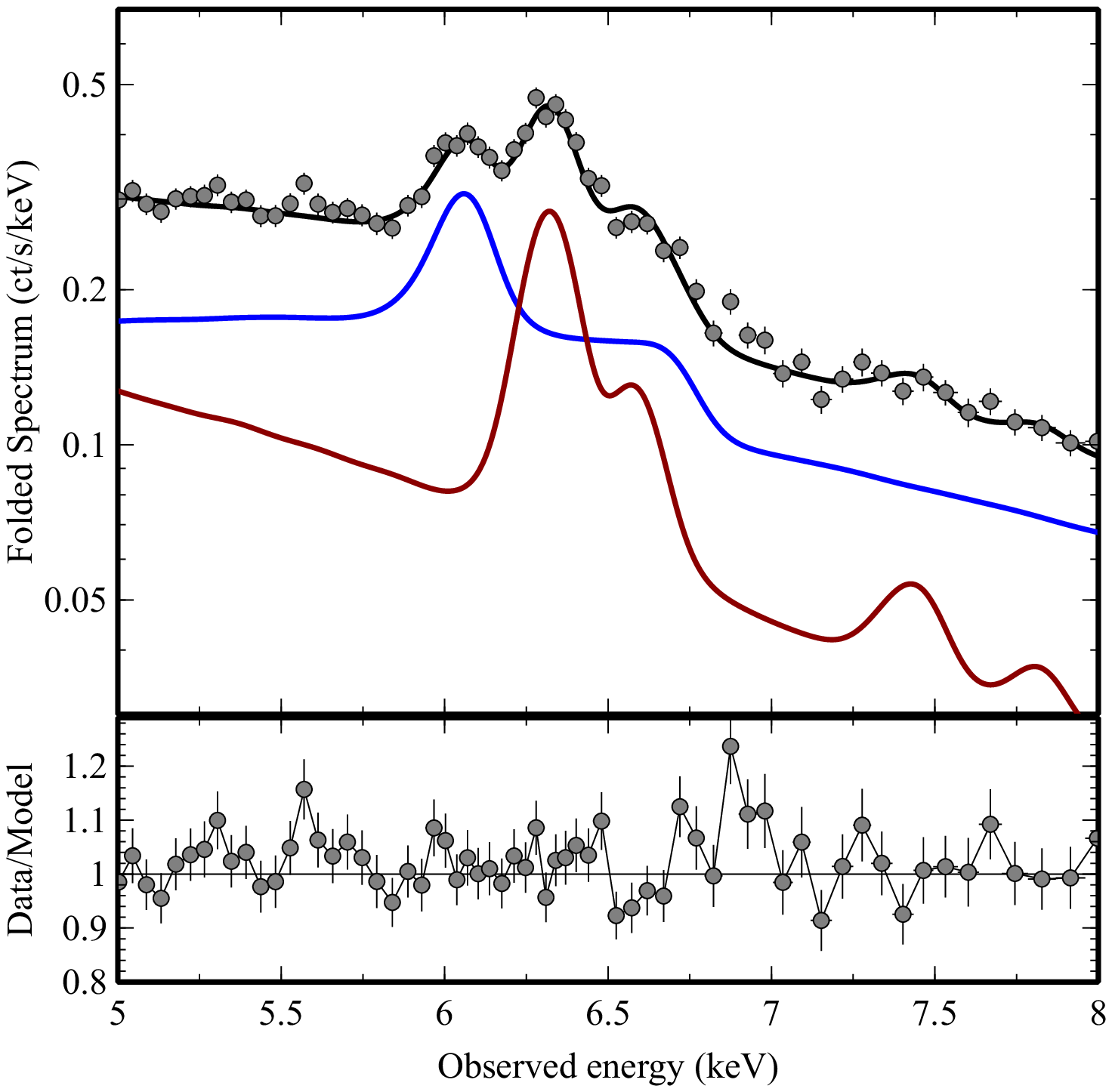,width=0.45\textwidth}
}
\caption{Fit of the ICM+cABS[PL+REFL] model to the {\it XMM-Newton}/EPIC-pn data.  {\it Left panel : }0.5--10\,keV folded EPIC-pn spectrum decomposed into contributions from the cool (red) and hot (brown) ICM components, plus the absorbed and reflected AGN power-law component (blue).  {\it Right panel : }Zoom-in on the iron-K region of the folded spectrum. The resolution of the EPIC-pn allows contributions to the iron complex from the fluorescent line of cold iron (6.40\,keV rest-frame; 6.06\,keV observed) as well as the radiative recombination lines of iron-25 and iron-26 (6.67\,keV and 6.97\,keV rest-frame; 6.32\.keV and 6.60\,keV observed) to be discerned.  The cold iron line is modeled as AGN reflection, whereas both ionized iron lines can be attributed to the hot component of the ICM.}
\label{fig:xmmspec}
\end{figure*}

We now present an analysis of the {\it XMM-Newton}/EPIC-pn spectrum of Cygnus~A.  A spatially resolved investigation of the interaction of the radio lobes with the ICM using these data has been presented previously by \cite{belsole:07a}.  However, to the best of our knowledge, the study presented here is the first published analysis of the {\it XMM-Newton} view of the Cygnus~A nucleus itself.  Based on previous works \citep{young:02a,arnaud:87a,ueno:94a}, we expect the X-ray spectrum within our extraction radius to be the superposition of thermal emission from the optically thin plasma of the ICM and a heavily absorbed AGN spectrum \cite[all absorbed by a Galactic column of $N_H=3.5\times 10^{21}\pcmsq$; ][]{dickey:90a}.    As we will see, the EPIC-pn derived constraints on the parameters of the AGN are rather poor due to the combination of the heavy absorption and the truncation of the EPIC bandpass at 10\,keV.  However, our main motivation here is to construct an ICM model that can be taken over into our {\it NuSTAR} analysis of the AGN.  

We find that the ICM emission is well described by two {\tt apec} components \citep{smith:01a} with independent temperatures ($T_1, T_2$).   Driven by physical considerations, we assume that the two ICM components have a common metal abundance ($Z$).  In this paper, all abundances are referenced to the solar abundance ($Z_\odot$) set described by \cite{wilms:00a}.   For all spectral models described below, the inclusion of a third temperature ICM component fails to give a significant improvement in the goodness of fit.   

We begin by approximating the AGN emission with a simple power-law (PL) continuum absorbed by a neutral/cold absorber at the redshift of Cygnus~A (cABS) with column density $N_H$ and solar abundances.  Both the intrinsic absorber and the Galactic absorbing column (which also affects the ICM emission) are describing using the {\tt tbabs} code of \cite{wilms:00a}.   This model  provides a good description of the broad-band spectrum (model ICM+cABS[PL] in Table~\ref{tab:xmmfits}).  The residuals do indicate, however, a strong unmodeled emission line between 6--7\,keV (rest-frame).   Adding a Gaussian emission line (emLINE) into the spectral model leads to a dramatic improvement in the goodness of fit ($\Delta\chi^2=-243$ for three additional model parameters).  The line is narrow ($\sigma<70\eV$ corresponding to a FWHM$<7700\kmps$), relatively strong (equivalent width $W=98\pm 12\eV$), and has a centroid energy precisely that expected from the fluorescent K$\alpha$ line of cold iron ($E=6.39\pm 0.01\keV$).  This model (ICM+cABS[PL+emLINE]) leaves no gross residuals in the fit.

In their {\it Chandra} study of Cygnus~A, \cite{young:02a} find soft power-law emission from a bipolar source spanning the nucleus that they identity as scattered AGN emission.  Hence, we must explore whether the ICM parameters can be skewed by the addition of such a scattered component.  Adding a power-law component that is not obscured by the AGN absorber with a photon index that is tied to that of the main AGN continuum but with a normalization decreased by a factor of 100 \citep{young:02a} has essentially no impact on the fit or the ICM parameters. Allowing the normalization to be free results in a small and insignificant improvement in the goodness of fit ($\Delta\chi^2=-7$ for 2 additional degrees of freedom) and a best fitting ICM abundance of $Z_{\rm Fe}=2.5^{+2.1}_{-0.9}\,Z_\odot$.  However, the normalization of this scattered component is then 25\% of the primary continuum --- this is 25$\times$ higher than found by {\it Chandra}, strongly suggesting that we should reject this possibility.   We conclude that soft/scattered emission from the AGN does not strongly bias our ICM model.

\begin{table*}[t]
\begin{center}
\begin{tabular}{lcc}\hline\hline
Spectral Model   & Parameters & $\chi^2/{\rm dof}$ \\\hline
ICM+cABS(PL)    &  $kT_1=1.69^{+0.20}_{-0.10}\keV, kT_2=6.74^{+0.69}_{-0.38}\keV, Z_{\rm ICM}=1.33^{+0.12}_{-0.09}Z_\odot$ & 1712/1605\\
& $N_{\rm H}=41.2^{+3.9}_{-3.0}\times 10^{22} \pcmsq, \Gamma=1.66^{+0.13}_{-0.11}$ \\\hline
ICM+cABS(PL+emLINE) & $kT_1=1.70^{+0.18}_{-0.07}\keV, kT_2=7.04^{+0.47}_{-0.45}\keV, Z_{\rm ICM}=1.40^{+0.11}_{-0.10}Z_\odot$   & 1469/1602\\
& $N_{\rm H}=34.1^{+3.0}_{-2.9}\times 10^{22} \pcmsq, \Gamma=1.43\pm 0.11$ & \\
& $E_{\rm line}=6.39\pm 0.01\keV, \sigma_{\rm line}<70\eV, W_{\rm line}=98\pm 12\eV$  & \\\hline
ICM+cABS(PL+REFL) & $kT_1=1.70^{+0.19}_{-0.09}\keV, kT_2=6.70^{+0.49}_{-0.38}\keV, Z_{\rm ICM}=1.30\pm 0.10Z_\odot$ & 1461/1603\\
& $N_{\rm H}=31.2^{+2.7}_{-2.5}\times 10^{22} \pcmsq, \Gamma=1.51^{+0.11}_{-0.10}$ & \\
& ${\cal R}=1.0^{+0.18}_{-0.37}, \log\xi<2.15, Z_{\rm refl}^f=Z_\odot$ \\\hline\hline
\end{tabular}
\end{center}
\caption{Spectral fits to our {\it XMM-Newton} EPIC-pn spectrum.  See text (\S\ref{sec:xmmanal}) for a detailed discussion of the spectral models.   All energies are quoted in the rest-frame of Cygnus~A (which has a cosmological redshift of $z=0.056$).  The ionization parameter $\xi_W$ is in units of $\ergcmps$.  Superscript {\it f} denotes a fixed parameter.  All errors are quoted at the 90\% level for one interesting parameter.  }
\label{tab:xmmfits}
\end{table*}

Physically, iron lines such as seen in our {\it XMM-Newton} spectrum are expected to arise from X-ray reflection by Compton-thick cold matter \citep{basko:78a,george:91a}; this matter can be identified with the outer regions of the accretion disk or a Compton-thick core to the obscuring torus. Hence, we replace the Gaussian line component with a cold X-ray reflection model as calculated by the {\tt xillver} code of \cite{garcia:13a} --- we set the abundance to solar but allow the ionization state of the reflector and the reflection fraction to be free parameters.   This model (ICM+cABS[PL+REFL]) provides a very comparable fit to the spectrum (Table~\ref{tab:xmmfits}).   The AGN parameters, especially those describing the X-ray reflection, are poorly constrained.  Allowing the metallicity of the reflector to be a free parameter fails to improve the fit and permits a strong degeneracy between the metallicity and the reflection fraction.   This is readily understood --- in the {\it XMM-Newton} band, the principle diagnostic of reflection is the energy and strength of the iron line, and a given line strength can be achieved by weak reflection from gas with high metallicity, or strong reflection from gas with low metallicity.  

Figure~\ref{fig:xmmspec} (left) shows the folded EPIC-pn spectral fits decomposed into its model components together with the fit ratios.  The ICM emission (consisting of $kT=1.7\keV$ and 6.7\,keV components) dominates the spectrum below 4\,keV, with the AGN emission becoming dominant at higher energies.  The iron line complex (Figure~\ref{fig:xmmspec} right) has distinct peaks corresponding to the 6.4\,keV fluorescent line of neutral iron (from X-ray reflection within the AGN), and the radiative-recombination lines of helium- and hydrogen-like iron at 6.67\,keV and 6.97\,keV respectively (from the hot ICM).  

\section{Analysis of the {\it NuSTAR} spectrum}

\begin{table*}
\begin{center}
\begin{tabular}{lcc}\hline\hline
Spectral Model & Parameters & $\chi^2/{\rm dof}$\\\hline

ICM+cABS(PL) &   $N_H=25.8^{+2.4}_{-2.2}\times 10^{22}\pcmsq, \Gamma=1.57\pm 0.02$  & 1410/1148 \\\hline

ICM+cABS(PL+emLINE) &  $N_H=19.8^{+2.3}_{-1.9}\times 10^{22}\pcmsq, \Gamma=1.54\pm 0.02$  & 1232/1146 \\
&  $E_{\rm emis}=6.34^{+0.03}_{-0.04}\keV, \sigma_{\rm emis}=0.01\keV^f, W_{\rm emis}=147^{+14}_{-25}\eV$  & \\\hline

ICM+cABS(PL+REFL) & $N_H=16.9^{+1.7}_{-1.1}\times 10^{22}\pcmsq, \Gamma=1.77\pm 0.03$  & 1237/1145\\
&   $\log\xi_{\rm refl}=2.7^{+0.1}_{-0.7}, {\cal R}=0.50^{+0.10}_{-0.05}, Z_{\rm refl}=2.3^{+0.8}_{-0.4} $  & \\\hline

ICM+cABS(PL+REFL+absLINE) & $N_H=22.4\pm 0.7\times 10^{22}\pcmsq, \Gamma=1.76^{+0.03}_{-0.05}$  & 1216/1143\\
&   $\log\xi_{\rm refl}<1.1, {\cal R}=0.43\pm 0.04, Z_{\rm refl}=2.5^{+0.6}_{-0.4} $  & \\
& $E_{\rm abs}=7.18\pm 0.08 \keV, \sigma_{\rm abs}^f=0.01\keV, W_{\rm abs}=-50^{+15}_{-18}\eV$  &  \\\hline

ICM+cABS(PL+REFL+emLINE) & $N_H=19.6^{+1.4}_{-2.3}\times 10^{22}\pcmsq, \Gamma=1.67^{+0.03}_{-0.05}$  & 1188/1143\\
&   $\log\xi_{\rm refl}=2.7^{+0.1}_{-0.3}, {\cal R}=0.40^{+0.07}_{-0.12}, Z_{\rm refl}=0.5^{+0.16}_{-0p} $  & \\
& $E_{\rm emis}=6.30\pm 0.04 \keV, \sigma_{\rm emis}^f=0.01\keV, W_{\rm emis}=110^{+24}_{-36}\eV$  &  \\\hline

ICM+cABS(PL+REFL+emLINE+absLINE) & $N_H=19.6^{+1.7}_{-2.4}\times 10^{22}\pcmsq, \Gamma=1.67^{+0.03}_{-0.05}$  & 1172/1141\\
&   $\log\xi_{\rm refl}=2.7^{+0.1}_{-0.3}, {\cal R}=0.39^{+0.05}_{-0.14}, Z_{\rm refl}=0.5^{+0.21}_{-0p} $  & \\
& $E_{\rm emis}=6.30\pm 0.04 \keV, \sigma_{\rm emis}^f=0.01\keV, W_{\rm emis}=107^{+27}_{-34}\eV$  &  \\
&  $E_{\rm abs}=7.24\pm 0.10 \keV, \sigma_{\rm abs}^f=0.01\keV, W_{\rm abs}=-35^{+17}_{-13}\eV$  &  \\ \hline

ICM+cABS*wABS(PL+wEMIS+REFL) &  $N_H=17.6\pm 2.3\times 10^{22}\pcmsq, \Gamma=1.60^{+0.04}_{-0.07}$ & 1163/1140\\
& $N_{Wabs}>3.4\times 10^{23}\pcmsq, \log\xi_W=3.4^{+0.5}_{-0.2} , v_{\rm Wabs}=-1.9^{+1.0}_{-0.7}\times 10^4\kmps$ & \\
& $N_{Wemis}=2.9^{+6.6}_{-1.9}\times 10^{23}\pcmsq, v_{\rm Wemis}=+3.2\pm 0.3\times 10^4\kmps$ & \\
& $\log\xi_{\rm refl}=2.7^{+0.1}_{-0.6}, {\cal R}=0.36^{+0.15}_{-0.11}, Z_{\rm refl}<1.3 $  &  \\\hline

ICM+cABS*2wABS(PL+wEMIS+REFL) &  $N_H=16.9\pm 1.4 \times 10^{22}\pcmsq, \Gamma=1.47^{+0.13}_{-0.06}$ & 1144/1137\\
& $N_{Wabs1}>3.0\times 10^{23}\pcmsq, \log\xi_{W1}=3.4^{+0.3}_{-0.2} , v_{\rm Wabs1}=-2.0^{+0.5}_{-0.6}\times 10^4\kmps$ & \\
& $N_{Wabs2}>3.6\times10^{22}\pcmsq, \log\xi_{W2}=3.0^{+1.3}_{-0.6} , v_{\rm Wabs2}=-5.0^{+3.0}_{-5.0}\times 10^4\kmps$ & \\
& $N_{Wemis}=5.0^{+15.0}_{-2.8}\times 10^{23}\pcmsq, v_{\rm Wemis}=+3.1\pm 0.2\times 10^4\kmps$ & \\
& $\log\xi_{\rm refl}=3.6^{+1.1}_{-0.6}, {\cal R}=0.71^{+0.9}_{-0.6}, Z_{\rm refl}<3.0 $  &  \\
\hline\hline
\end{tabular}
\end{center}
\caption{Spectral fits to the 3--70\,keV {\it NuSTAR} data.  See text (\S\ref{sec:nustar_basic} and \S\ref{sec:nustar_wind}) for a detailed discussion of the spectral models. All energies are quoted in the rest-frame of Cygnus~A (which has a cosmological redshift of $z=0.056$).  All ionization parameters ($\xi_W, \xi_{W1}, \xi_{W2}, \xi_{\rm refl}$) are in units of $\ergcmps$.   Superscript {\it f} denotes a fixed parameter, and sub-script {\it p} denotes a parameter that has hit the upper/low limit of its allowed range.   All errors are quoted at the 90\% level for one interesting parameter. }
\label{tab:nustarfits}
\end{table*}

\subsection{Basic Characterization : The Detection of Compton Reflection}
\label{sec:nustar_basic}

\begin{figure}[t]
\centerline{
\psfig{figure=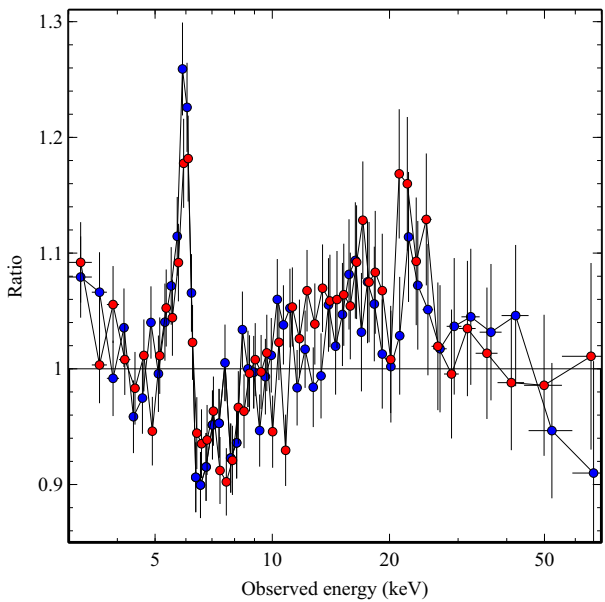,width=0.45\textwidth}
}
\caption{Ratio of the {\it NuSTAR} FPMA (red) and FRMB (blue) data to a spectral model in which the AGN is described by a simple absorbed power-law (model ICM+cABS[PL] in Table~\ref{tab:nustarfits}).  The spectral model includes a two-temperature ICM component with shape (but not normalization) fixed to that found in the {\it XMM-Newton} analysis.  }
\label{fig:nustar_xillver_ratio}
\end{figure}

\begin{figure*}[t]
\hbox{
\psfig{figure=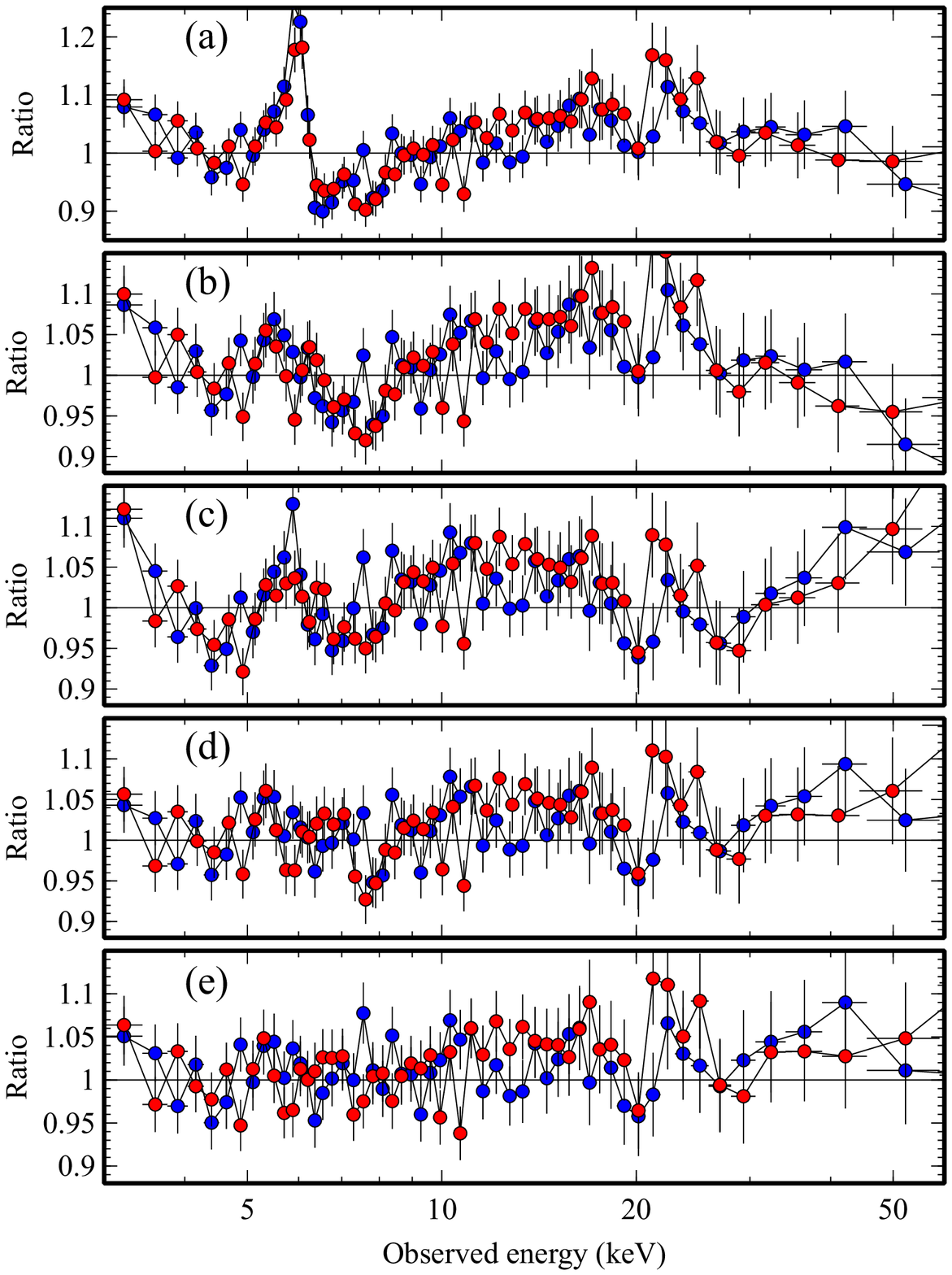,width=0.48\textwidth}
\psfig{figure=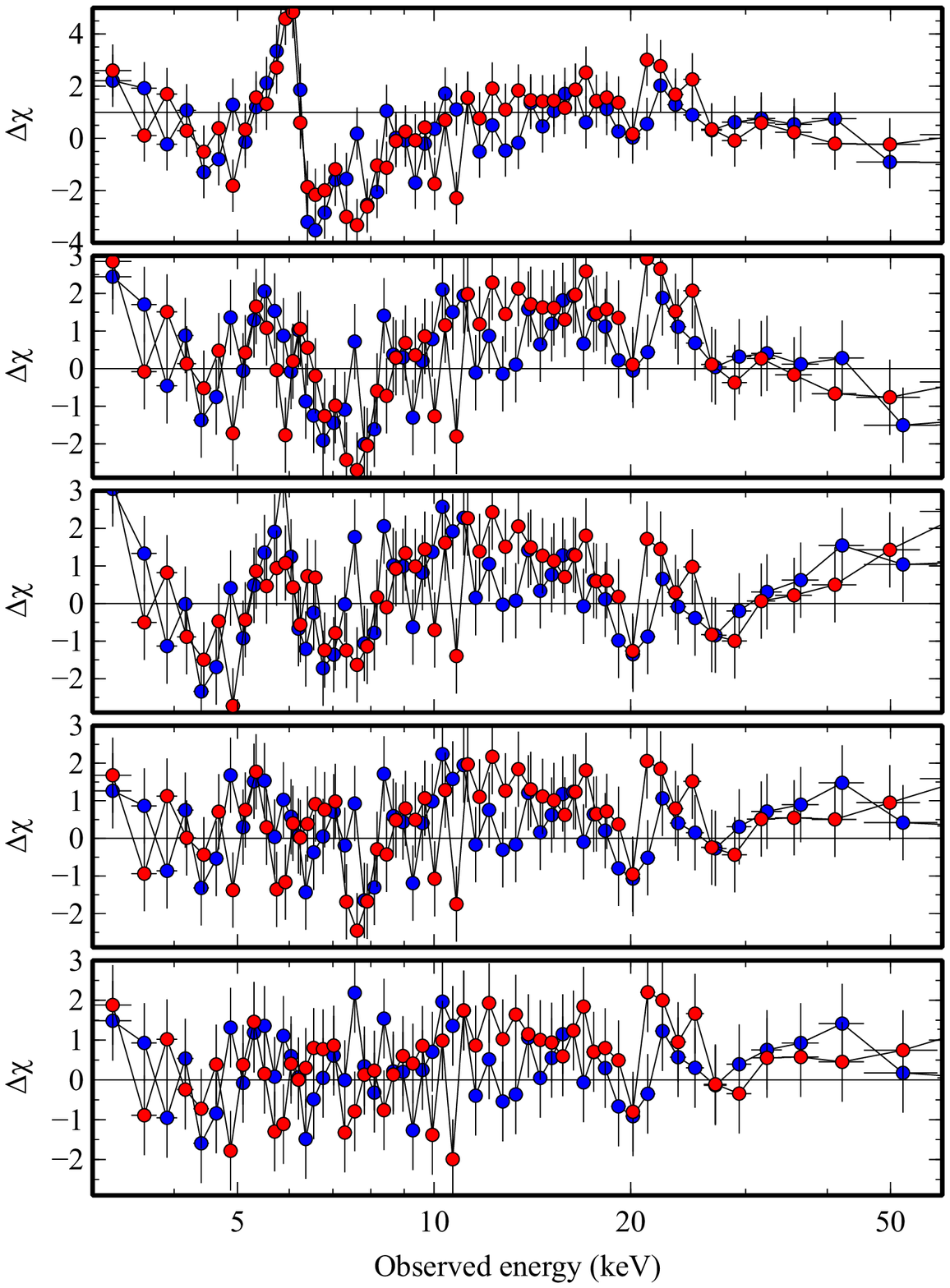,width=0.48\textwidth}
}
\caption{Data-to-model ratios (left) and $\Delta\chi$ (right) resulting from our spectral analysis of the {\it NuSTAR} data (FPMA=red; FPMB=blue).   In all cases, the spectral model includes a two-temperature ICM component with shape (but not normalization) fixed to that found in the {\it XMM-Newton} analysis.  The rows differ in the description of the AGN spectrum.    {\it Row-(a) :  }AGN described as a power-law absorbed by a cold column [cABS(PL)].   {\it Row-(b) : }AGN described by absorbed power-law plus narrow emission line [cABS(PL+emLINE)].  {\it Row-(c) : }AGN described by absorbed power-law with a reflection component [cABS(PL+REFL)].   {\it Row-(d) : }AGN described by absorbed power-law with reflection, plus the joint absorption/emission signatures of a single-component fast highly ionized wind.  {\it Row-(e) : }Same as row-(d) except for the inclusion of a second ionized absorption component.   Error bars are 1-$\sigma$.}
\label{fig:nustar_specrat}
\end{figure*}

For an initial impression of the {\it NuSTAR} spectrum, we compare the data with a spectral model consisting of a two-component ICM and a power law continuum modified by intrinsic cold absorption.  The ``shape parameters'' of the ICM spectrum (i.e. the temperatures, the common abundance, and the relative normalizations of the two components) are fixed to those of our best-fitting model for the {\it XMM-Newton} data.  The total normalization of the ICM emission is allowed to depart from the {\it XMM-Newton} value to account for the different point spread functions and any possible flux cross-calibration issues.  This model is not a particularly good description of the data [model ICM+cABS(PL) in Table~\ref{tab:nustarfits}].  As can be seen in Fig.~\ref{fig:nustar_xillver_ratio} (also Fig.~\ref{fig:nustar_specrat}a), this model leaves obvious residuals in the form of an emission line at iron-K$\alpha$ energies and a weak broad hump peaking at 20--30\,keV.  This high-energy hump is suggestive of Compton reflection from optically thick matter.

Adding a narrow Gaussian emission line to the model leads to a dramatic improvement in the goodness of fit ($\Delta\chi^2=-178$ for two additional model parameters; model ICM+cABS[PL+emLINE] in Table~\ref{tab:nustarfits}).  As expected, the putative Compton reflection hump is still not captured by this model, leading to the subtle high-energy hump visible in panel-b of Fig.~\ref{fig:nustar_specrat}.  Of more interest is the fact that the simple emission line model leaves significant unmodeled complexities in the iron-K band (6--8\,keV), and that the best-fitting energy of the line (rest-frame $E=6.34^{+0.03}_{-0.04}\keV$ at the 90\% confidence level) is inconsistent with that expected from any charge-state of iron.  While it may be tempting to interpret this as the effects of gravitational redshifting of an accretion disk reflection spectrum, detailed spectral modeling rejects this possibility.  The addition of a relativistic disk reflection component \citep[modeled with the {\tt relxill\_lp} code of][]{garcia:14a} leads to no significant improvement in the goodness of fit ($\Delta\chi^2=-1$ for 5 additional model parameters), even if we permit the height of the irradiating source, and the metallicity, ionization state, inclination and inner edge of the accretion disk to all be free parameters.  

This iron-K band complexity is {\it not} simply due to the presence of an iron-edge in reflection.   Adding a Compton reflection component to describe reflection from the outer accretion disk or obscuring torus (REFL; see Section~\ref{sec:xmmanal})  improves the fit at 10--30\,keV but does not account for the spectral complexity in the iron-K band.  Indeed, since the reflection component forces the iron-K$\alpha$ line to be at 6.4\,keV, the overall fit is slightly worse than the more phenomenological ICM+cABS[PL+emLINE] model.   The iron-K band residuals persist even if the ionization state and metallicity of the reflector are allowed to be free parameters (model ICM+cABS[PL+REFL] in Table~\ref{tab:nustarfits}; also see panel-c of Fig.~\ref{fig:nustar_specrat}).  Since this reflection model now includes a Compton hump, the underlying power-law continuum is inferred to be steeper (photon index $\Gamma=1.77\pm 0.03$ compared with $\Gamma=1.54\pm 0.02$ for models without the Compton reflection).  However, the steepening of the primary continuum within this reflection model now leaves high-energy residuals in the form of a hard tail above 40\,keV (Fig.~\ref{fig:nustar_specrat}, panel-c).

Still, in gross terms, this simple model (ICM emission plus an absorbed AGN power-law continuum with associated reflection) provides a good description of the {\it NuSTAR} spectrum.  The presence of Compton-hump clearly points to the presence of scattering from a Compton-thick, or near Compton-thick, structure in the central engine of Cygnus~A.  We return to this issue in Section~\ref{discussion} --- here we simply point out that this matter must lie outside of our line-of-sight since our same spectrum constrains the absorption column to the central X-ray source to be only $N_H=1.69^{+0.17}_{-0.11}\times 10^{23}\pcmsq$ (approximately 0.12 Thomson depths).  To the best of our knowledge, this is the first direct demonstration of Compton-thick circumnuclear material in the nucleus of Cygnus~A.   

\subsection{Modeling the Detailed Iron-K Band Structure : Detection of a Fast Ionized Wind}
\label{sec:nustar_wind}

We now return to the iron-K band complexity noted above.  Referenced to the fiducial simple reflection model, the 4--8\,keV spectrum shows correlated sets of residuals at the $\pm 5\%$ level, with excursions out to $\pm10\%$ level (Fig.~\ref{fig:nustar_specrat}, panel-c).  The correlated nature of these residuals argues against simple Poisson fluctuations (which, for the binning used in Fig.~\ref{fig:nustar_specrat}, are $\pm 3$\%).  Furthermore, these residuals are at a substantially greater level than expected effective-area calibration errors which should be at the $<1\%$ level based on an analysis of {\it NuSTAR} spectra of the Crab nebula matched to the same off-axis angles \citep[also see ][]{madsen:15a}.  

Given that Cygnus~A has strong ICM emission lines, we also need to consider the effects of uncertainties in our description of the ICM as well as residuals arising from uncertainties in the energy calibration of the FPMs.  Allowing the shape-parameters of the two-temperature ICM model to vary within the 90\% error ranges determined by {\it XMM-Newton} has a negligible effect on the fit and the residuals ($\Delta\chi^2<0.1$ for three additional but constrained degrees of freedom).  If we allow the ICM model to be completely free, the normalization of the low-temperature component increases by a factor of 100.  The corresponding impact on the iron-K band, driven by the very strong increase of the 6.7\,keV line complex, has a significant impact on the residuals under discussion here.  However, this now brings the ICM model into very strong conflict with {\it XMM-Newton}, over-predicting the soft ($<2\keV$) emission by a fact of 5--10.  We conclude that ICM models that are compatible with the {\it XMM-Newton} data leave these residuals in the {\it NuSTAR} fit.    To address the calibration question, we refit our fiducial reflection model permitting a free gain-scale --- this can reduce the magnitude of the iron-K band residuals if there is an energy-scale offset in the iron-K band of 120\,eV (in both FPM).   However, the gain scale of {\it NuSTAR} has been calibrated on a pixel-by-pixel basis using the onboard radioactive calibration sources.  The expected residual gain calibration error this early in the mission is $\delta E<20\eV$ \citep{madsen:15a}.  These considerations suggest that the residuals are real and have an astrophysical origin. 

The pattern of residuals suggests that, in addition to the spectral features associated with Compton-thick X-ray reflection, there are both additional emission and absorption components.  Given that relativistic disk reflection fails to explain these features, we turn our attention to wind models.  Specifically, we suggest that these spectral features correspond to blue-shifted iron-K absorption together with (redshifted) iron-K emission from the back-side of the wind (i.e., an iron-K band P-Cygni profile).

As a first exploration of this wind hypothesis, we add a narrow Gaussian emission line (emLINE) and/or a narrow absorption line (absLINE) to our base-line model ICM+cABS(PL+REFL).  These three additional model fits are presented in Table~\ref{tab:nustarfits}.  Adding just the absorption line leads to a significant improvement in the goodness of fit ($\Delta\chi^2=-21$ for two additional model parameters) but still leaves significant iron-band residuals.  The addition of just the extra emission line produces a much greater improvement in the goodness of fit ($\Delta\chi^2=-59$ for two additional model parameters).   However, the residuals left by this model still indicate the presence of absorption, and so it is not surprising the best model (ICM+cABS[PL+REFL+emLINE+absLINE]) incorporates both emission and absorption lines.

A more physical description of the wind requires photoionization modeling.   We use the XSTAR code \citep{kallman:01a} to compute both the X-ray absorption and emission spectra from a photo-ionized slab of matter (with solar abundances) as a function of ionization parameter $\xi_W$ and column density $N_W$.  Our grid of models cover the range $\xi_W\in(10,10^5)\ergcmps$ in logarithmic steps of $\Delta \xi_W/\xi_W=0.2$ \citep[sufficient to resolve the structures in ionization space; ][]{reynolds:12a}, and the range $N_W\in(10^{21},10^{25})\pcmsq$ with steps $\Delta N_W/N_W=0.2$.     Guided by the simple-line fitting results above, we expect velocities in the $0.03-0.1c$ range.  Thus, we construct the XSTAR models assuming line-of-sight velocity spreads of $10,000\kmps$ (formally, this is achieved by setting the turbulent velocity parameter to $10,000\kmps$).  We note that the results described below do not change appreciably if we adopt smaller velocity spreads. 

\begin{figure*}[t]
\centerline{
\psfig{figure=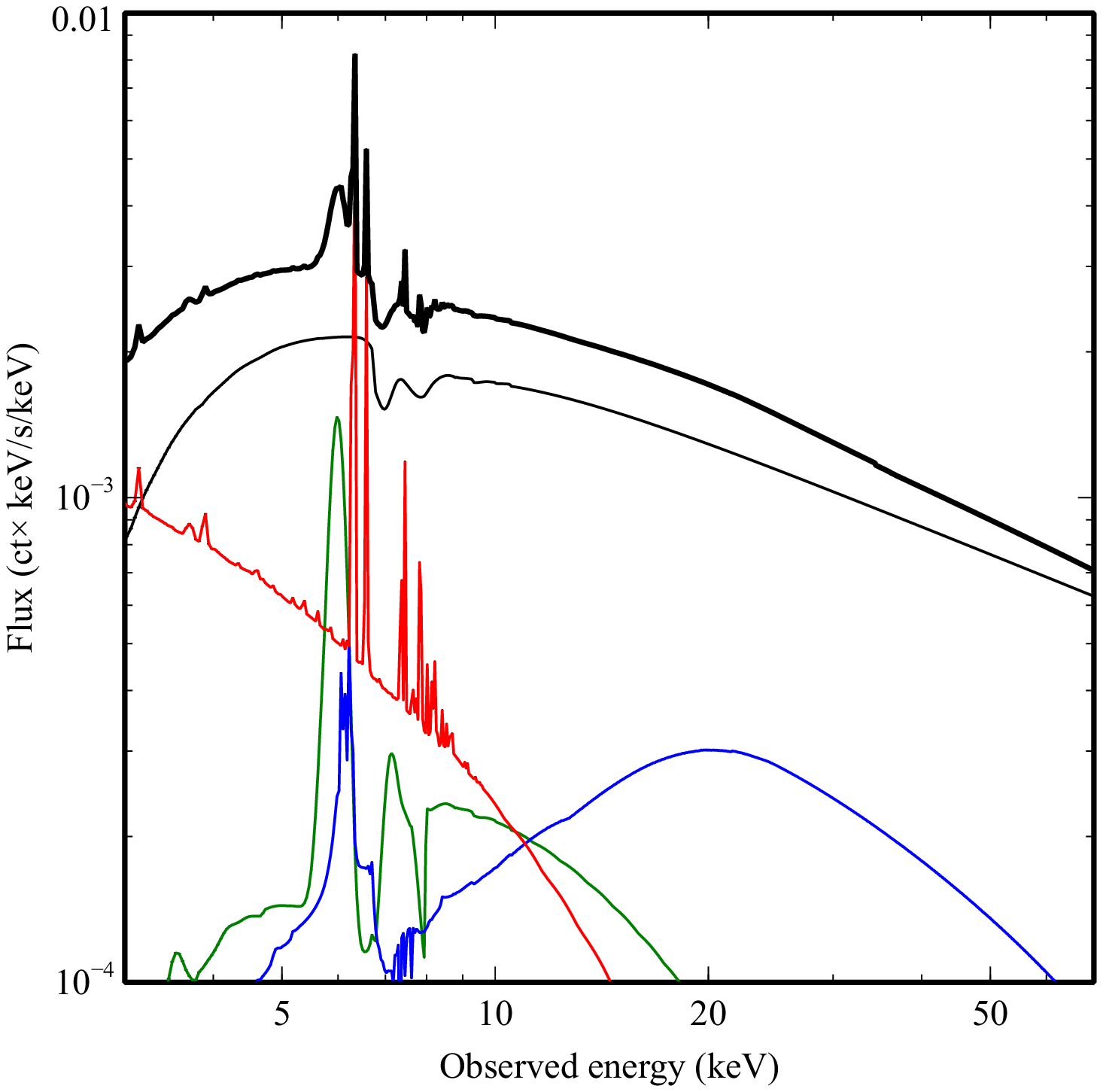,width=0.45\textwidth}
\hspace{0.5cm}
\psfig{figure=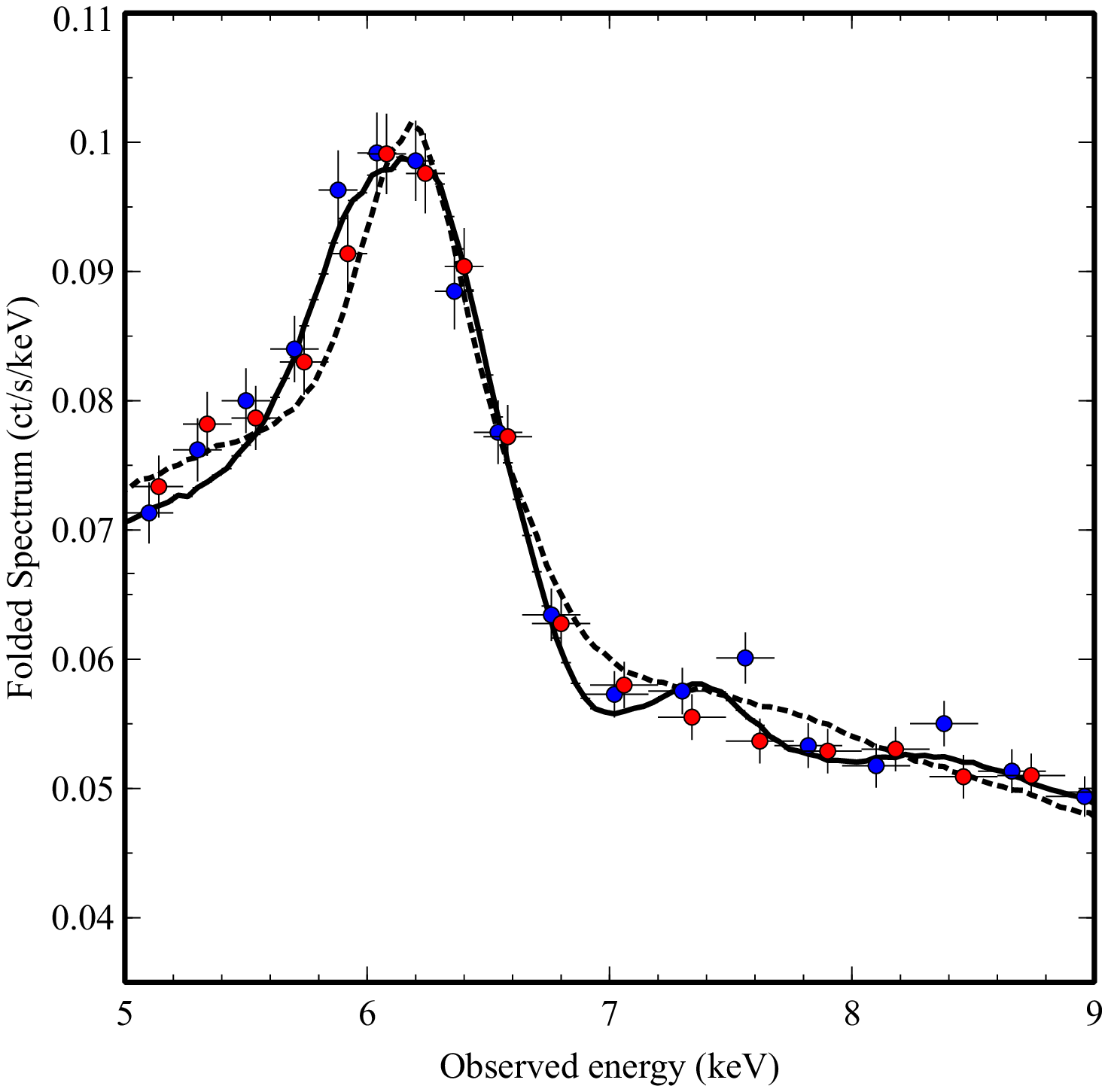,width=0.45\textwidth}
}
\caption{{\it Left panel : } Our best fitting {\it NuSTAR} spectral model for Cygnus~A.  Shown here are the power-law continuum absorbed by both a neutral column and a highly ionized fast outflow (thin black line), the hot ICM component (red line), reflection from ionized Compton-thick matter (blue line), and the redshifted re-emission from the ionized wind (green line).  {\it Right panel : }    Folded {\it NuSTAR} spectrum zoomed in on the iron complex.  The FPMA and FPMB data have been co-added and binned for plotting purposes (but not for fitting).   The solid and dashed lines show the best model fits with and without the fast ionized wind.}
\label{fig:nustar_ironline}
\end{figure*}

Augmenting our base-line spectral model for Cygnus~A (ICM+cABS[PL+REFL]) with a blueshifted absorption component (wABS) and a redshifted emission component (wEMIS) of the same ionization parameter $\xi_W$ produces a significant improvement in the goodness of fit, $\Delta\chi^2=-74$ for 5 additional model components.   The iron band residuals are, to a very large degree, described by these wind features (Fig.~\ref{fig:nustar_specrat}, panel-d).  Table~\ref{tab:nustarfits} reports the parameter values for this model.  We find that, formally, we only obtain a lower-limit on the column density of the absorber, $N_{Wabs}>3.4\times 10^{23}\pcmsq$, and that we require the wind to be highly ionized, $\log\xi_W=3.2-3.9$.  We note that the XSTAR model, as constructed, does not correctly capture Compton scattering and hence does not formally reject Compton-thick solutions.    Confirming expectations from the simple line fitting, the velocity of the absorbing outflow is in the range $12,000-29,000\kmps$.  The emission component of the outflow has a fitted column density that is comparable with, or maybe slightly less than, the absorption component ($N_{\rm Wemis}=1.0-10\times 10^{23}\pcmsq$).   This shows that the outflow extends a significant solid angle as seen from the central X-ray source. 

This description, with a single ionized absorption component, still leaves a weak unmodeled absorption feature in the residuals (at 7.5\,keV observed energy; Fig.~\ref{fig:nustar_specrat}d).  Adding a second ionized absorption component (wABS) with its own independent column density, ionization parameter, and velocity, leads to a further improvement in the fit ($\Delta\chi^2=-19$ for three additional model parameters).  While we have concluded the possible presence of this second wind component from systematic spectral analysis, we note that the presence of (two) absorption lines at observed energies of 6.5\,keV and 7.5\,keV is suggested even by examining the residuals from a fit of a simple absorbed power-law to the data (Fig.~\ref{fig:nustar_xillver_ratio}).   We note however, while it appears to be very fast ($20,000-100,000\kmps$), this second wind component has poorly constrained properties.   The full spectral model for this two-component ionized wind model is shown in Fig.~\ref{fig:nustar_ironline}, together with a zoom-in on the iron line region of the spectrum.

\begin{figure*}[t]
\centerline{
\psfig{figure=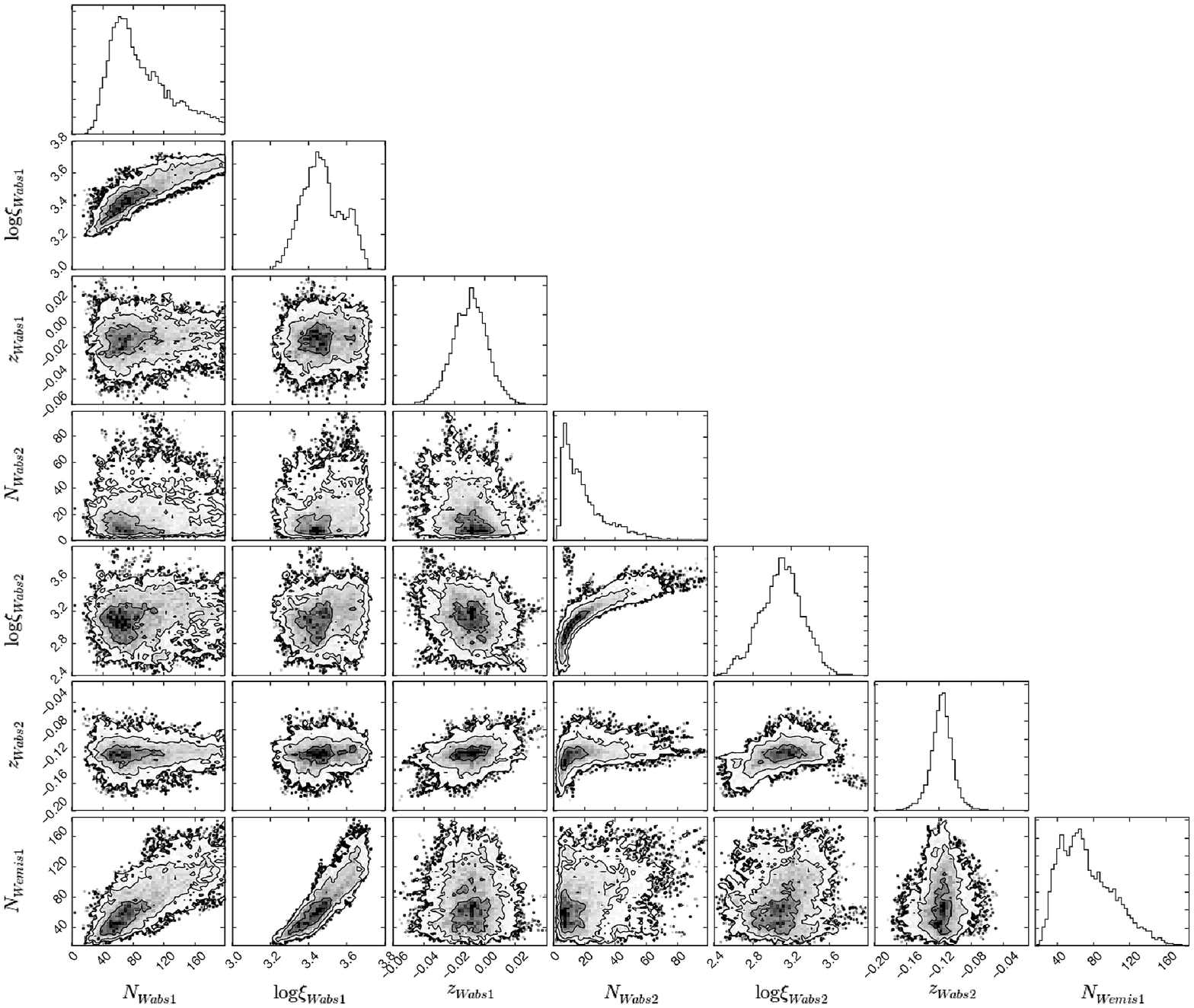,width=0.85\textwidth,angle=270}
}
\caption{Results of MCMC analysis of the wind model as applied to the {\it NuSTAR} data.  The column densities of the absorbing ($N_{Wabs}$) and emitting ($N_{Wemis}$) components of the wind are in units of $10^{23}\pcmsq$.  The redshift of the absorbing component ($z_{Wabs}$) is with respect to the observer, and so $z_{Wabs}=0.056$ correspond to being at rest with respect to the nucleus of Cygnus-A.  $\Gamma$ is the photon index of the primary continuum.   The histograms on the left-hand edge of each row show the probability distribution for the parameter in question and are normalized to unity. }
\vspace{0.5cm}
\label{fig:triangle}
\end{figure*}

To a large degree, the poor constraints from the wind fit result from strong covariances between model parameters.  To uncover these covariances, we have performed a Goodman-Weare Monte Carlo Markov Chain (MCMC) analysis of our two-component wind model as implemented in Jeremy Sanders' {XSPEC\_EMCEE} package.  The Goodman-Weare algorithm \citep{foreman:12a} simultaneously steps a set of $N$ walkers through parameter space, with each walker taking a random step in the direction of another (distinct) randomly chosen walker.  In addition to having excellent convergence properties, this flavor of MCMC does not require a previously defined ``proposal'' (i.e. probability distributions detailing how each MCMC step should be taken) and is superb for uncovering previously unknown covariances.  We run 50 walkers for 5000 steps (leading to a 250,000 element chain) with a 1000 step burn-in period.   Since our XSTAR models do not accurately capture the effects of Compton scattering, we impose a prior that the column density not exceed $N_W=3\times 10^{24}\pcmsq$ (i.e. approximately two Compton depths).  While we do not exclude the possibility of higher column densities, our current models certainly will be a poor description of such winds.   

The results of the MCMC analysis for selected parameters are shown in Fig.~\ref{fig:triangle}.  Several interesting covariances are found.   We see that the column density of the wind components (both absorption and emission) are strongly positively correlated with their ionization parameter.  This is readily understood, at least for the absorber, when it is realized that the data are sensitive to essentially only one strong absorption line (FeXXVI-K$\alpha$).   For sufficiently high ionization parameters, the fraction of iron in that charge state is $f_{FeXXVI}\propto \xi_W^{-1}$, and so the column density of that ion is $N_{FeXXVI}\propto N_{Wabs}\xi_W^{-1}$.  Given that observations fix the depth of the absorption line and hence the column density of FeXXVI, we expect a degeneracy of the form $N_{Wabs}\propto \xi_W$. Similar arguments drive the degeneracy between $N_{Wemis}$ and $\xi_W$.  Of course, an immediate consequence of these degeneracies with $\xi_W$ is that the two wind column densities are also covariant.     We note that the wind velocity is free from any obvious degeneracies/covariances.  

\subsection{Constraints on the High-Energy Continuum Cutoff}

If the X-ray continuum is formed by thermal Comptonization in an accretion disk corona with electron temperature $T_e$, we expect the power-law spectrum to possess an exponential cutoff with a characteristic energy of $E_{\rm cut}\approx 2-3kT_e$.   Adding an exponential cutoff to the continuum component of our best-fitting wind$+$reflection spectral model and refitting produces a slight improvement in the goodness of fit ($\Delta\chi^2=-4.1$ for one additional model parameter).   A blind application of the F-test would suggest that the cutoff is significant at the 90\% level (but not the 95\% level).  However, the F-test is not well posed in this situation since the null model (with no cutoff) is not fully embedded in the expanded model (with a cutoff), existing only in the $E_{\rm cut}\rightarrow \infty$ limit \citep{protassov:02a}. Thus, we conclude that we cannot claim the detection of a cutoff.  The  lower-limit on the cutoff energy is $E_{\rm cut}>111\keV$ at 90\% confidence ($>101\keV$ at 99\% confidence).  We note that Compton-reflection is strongly required even when a continuum cutoff is included in the spectral model --- including the cutoff in our single-wind model does not lead to a statistically-significant change in the require Compton reflection fraction (shifting the 90\% error range from ${\cal R}=0.25-0.51$ to ${\cal R}=0.13-0.35$).

\subsection{Do We See the Ionized Wind in Previous X-ray Data?}

With the 2013 {\it NuSTAR} data suggesting the existence of a high-column density, high-ionization, and fast wind, we must return to the {\it XMM-Newton}/EPIC-pn dataset (\S4) and ask whether these wind signatures are also present during this 2005 observation.  We can also ask whether hints of these signatures have been seen in any previous X-ray observations.  

Applying the photoionization wind model, we find that the 2005 EPIC-pn spectrum requires neither the blueshifted absorption nor the redshifted emission from a fast wind.  With an assumed line-of-sight velocity spread of $10,000\kmps$ and fixing the ionization parameter and velocity at the best-fitting {\it NuSTAR} values, the formal limits on the column density of the wind are $N_{Wabs}<1.2\times 10^{23}\pcmsq$ and $N_{Wemis}<4.5\times 10^{22}\pcmsq$ (90\% confidence), strongly inconsistent with the {\it NuSTAR} values.  

Thus, we must appeal to time-variability in order to reconcile these two datasets.  In fact, time-variability is clearly required irrespective of the detailed spectral modeling --- simple PL+emLINE characterization requires the line centroid to shift from $6.39\pm 0.01\keV$ in the {\it XMM-Newton} data to $6.34^{+0.03}_{-0.04}\keV$ in {\it NuSTAR} \citep[a change that exceeds the 20--30\,eV gain calibration of {\it NuSTAR}; ][]{madsen:15a}.  

We suggest that the fast ionized wind has appeared during the 7.5 years between these observations.  In the same period, the 2--10\,keV (de-absorbed) luminosity of the AGN has more than doubled, changing from $1.7\times 10^{44}\ergps$ during the {\it XMM-Newton} observation to $3.7\times 10^{44}\ergps$ during our {\it NuSTAR} pointing \footnote{The observed/absorbed 2--10\,keV AGN luminosity increased from $9\times 10^{43}\ergps$ during the {\it XMM-Newton} observation to $2.0\times 10^{44}\ergps$ during the {\it NuSTAR} pointing.  }.    Over the same time-frame, the cold/neutral intrinsic column has {\it dropped} from $N_H\approx 3.1\times 10^{23}\pcmsq$ to $N_H\approx 1.4\times 10^{23}\pcmsq$.  Thus, it is tempting to suggest that some fraction of the ionized outflowing matter seen by {\it NuSTAR} originated from the cold material that obscured the AGN during the {\it XMM-Newton} observation, possibly in (highly non-linear) response to an increase in the continuum luminosity.

What about other X-ray datasets?  {\it Suzaku} observed Cygnus~A on 15-Nov-2008 with 46\,ks of good on-source exposure.  The ICM aspects of this dataset have been previously discussed by \cite{sarazin:12a}.   To examine the core, we have extracted the spectra from the three operating X-ray Imaging Spectrometers (XISs) on {\it Suzaku} using an extraction radius of 3\arcmin.  Fitting our fiducial ICM+cABS[PO+REFL] model to the 0.7--10\,keV XIS spectra (with ICM parameters allowed to be free) we note residuals at the $\pm 10\%$ level in the 6--9\,keV region.  However, there are deviations of the front- and back-illuminated XISs at the same level, and there is an obvious effect from a background line at 7.5\,keV.  We conclude that the {\it Suzaku}-XIS data can neither confirm or refute the presence of wind-like spectral residuals.  

The other observatory capable of, in principle, witnessing the spectral signatures of a highly ionized wind is {\it Chandra}.  In the vast majority of {\it Chandra}/ACIS data on Cygnus~A in the archive, studies of the AGN spectrum are strongly compromised by severe photon pile-up.  However, there is a small amount of exposure time (9\,ks) taken in a short frame-time (0.4\,s) mode on 26-May-2000.  A detailed study of the AGN using these data has been conducted by \cite{young:02a}.   They note the existence of an anomalous absorption feature blue-wards of the 6.4\,keV fluorescent emission line which they empirically describe using an absorption edge with a (rest-frame) threshold energy of $7.2\pm 0.1$\,keV \citep[see Fig.~5 of ][]{young:02a}.  While further study of this feature was not possible in such a short exposure, it is tempting to suggest that this feature resulted from iron-K$\alpha$ line absorption in an ionized wind.  We note that, as was the case during our {\it NuSTAR} observation, the obscuring column to the AGN central engine during this {\it Chandra} observation was lower than that found during the {\it XMM-Newton} pointing.

\begin{table}
\begin{center}
\begin{tabular}{lc}\hline\hline
Observed  & $F_{3-10\keV}=3.1\times 10^{-11}\ergpcmsqps$ \\
Fluxes & $F_{10-30\keV}=5.6\times 10^{-11}\ergpcmsqps$ \\
 & $F_{30-80\keV}=6.8\times 10^{-11}\ergpcmsqps$ \\\hline
 Intrinsic AGN  & $L_{2-10\keV}=3.7\times 10^{44}\ergps$\\
 luminosities& $L_{10-30\keV}=4.2\times 10^{44}\ergps$\\
 & $L_{30-80\keV}=5.1\times 10^{44}\ergps$\\\hline\hline
\end{tabular}
\end{center}
\caption{Observed fluxes (including ICM contributions and effects of absorption), and intrinsic AGN luminosities (excluding ICM and de-absorbing), based upon our best-fitting model from Table~\ref{tab:nustarfits}}
\label{tab:fluxesandlums}
\end{table}

\section{Discussion and Conclusions}
\label{discussion}

\subsection{Compton Scattering in Circumnuclear Material}

{\it NuSTAR} has given us the cleanest view yet of the nuclear hard X-ray emission of Cygnus~A.  The source-dominated spectrum across the 3--70\,keV band permits us to detect spectral curvature consistent with Compton scattering from cold, Compton-thick (or near Compton-thick) material that may be identifiable with the outer regions of the accretion disk or a high column density ``core'' of the obscuring torus of unified AGN schemes.  In the spectral modeling presented above, the Compton scattering and associated iron fluorescence was described using the {\tt xillver} model of \cite{garcia:13a} --- we stress that, while it is usually employed to describe reflection from accretion disks, we are using {\tt xillver} in the spirit of a generic, self-consistent, and accurate model of X-ray reflection from Compton-thick matter.  

Another view of the Compton scattered component can be gained by using the {\tt MYTorus} model of \cite{murphy:09a}.  While possessing less up-to-date atomic physics data and a simpler (non self-consistent) ionization structure, this model allows us to explore scattering/reflecting distributions that are not completely Compton-thick.  Replacing the {\tt xillver} component with the scattering and iron line emission described by {\tt MYTorus} in our ultimate, best-fitting two-zone ionized wind model, we find that the fit is comparably good, with the column density of the scatterer constrained to be $N_{\rm H,scat}=1.9^{+1.0}_{-0.8}\times 10^{24}\pcmsq$ (0.7-2.0 Compton-depths) at 90\% confidence, and $N_{\rm H,scat}=1.9^{+2.3}_{-1.1}\times 10^{24}\pcmsq$ (0.5-2.8 Compton-depths) at 99\% confidence.  Taken at face value, this suggests that the scattering/reflecting matter is only mildly Compton-thick.   Confirming our previous expectations, the scattering/reflecting column density is substantially more than that seen in (cold) absorption, demonstrating that the scattering matter must be out of the line of sight. 

\subsection{The X-ray Loudness of Cygnus~A}

We use our best fitting (wind$+$reflection) model to derive intrinsic AGN luminosities, correcting for the contribution from the ICM and de-absorbing.  The results are shown in Table~\ref{tab:fluxesandlums}.  The 2--10\,keV X-ray luminosity is $3.7\times 10^{44}\ergps$, almost 10\% of the bolometric luminosity as calculated by \cite{privon:12a} on the basis of the radio-to-IR SED.  Unless the  IR-calorimetry assumption underlying the \cite{privon:12a} work fails due to some unusual geometry, we conclude that Cygnus~A is X-ray loud, with a 2-10\,keV bolometric correction of only 10.  However, given the fact that Cygnus~A likely has an Eddington ratio of only $10^{-2}$, this behavior falls in line with that found in other broad-line AGN \citep{vasudevan:09a}.

\subsection{Physical Properties of the Wind}

Our most novel finding in the {\it NuSTAR} spectrum is the presence of subtle features indicative of a high-column density ($N_{Wabs}>3\times 10^{23}\pcmsq$), highly ionized, ($\xi\sim 2500\ergcmps$), fast ($v=-2.0^{+0.5}_{-0.6}\times 10^4\kmps$) wind.  The fact that we see possible red-shifted emission from this wind suggests that it is a wide-angle outflow, subtending a significant solid-angle of the sky as seen from the central X-ray source.   Here, we discuss the inferred physical properties of this wind.

Suppose that the wind signatures originate from clouds with characteristic hydrogen number density $n$ and volume filling factor $f$ at a characteristic radius $r_0$.  The observed column density is $N_W=nr_0f$ and the ionization parameter is $\xi=L_i/n r_0^2=L_if/N_Wr_0$ ($L_i$ is the ionizing luminosity of the central source).  The mass flux is the wind is then given by
\begin{eqnarray}
\dot{M}_W&=&\Omega f m_p\mu r_0^2v_Wn \\
&=&\Omega m_p\mu N_Wr_0v_W,
\end{eqnarray}
where $\Omega$ is the solid angle subtended by the wind as viewed from the center of the system, $v_W$ is the velocity of the wind, and $\mu$ is the average particle mass of the wind per proton (in units of the proton mass $m_p$).   We now make the assumption that the wind is flowing at its local escape velocity so that the characteristic radius is 
\begin{equation}
r_0=2\left(\frac{c}{v_W}\right)^2r_g.  
\end{equation}
With this, we can write useful expressions for the mass flux
\begin{equation}\label{eq:massflux}
\dot{M}_W=2\Omega m_pc^2\mu r_g N_Wv_W^{-1},
\end{equation}
scalar momentum flux ($P_W$),
\begin{equation}\label{eq:mtm}
P_W=\dot{M}_Wv_W=2\Omega m_pc^2\mu r_g N_W,
\end{equation}
and kinetic energy flux ($L_K$), 
\begin{equation}\label{eq:energyflux}
L_K={1\over 2}\dot{M}_Wv_W^2=\Omega m_pc^2\mu r_g N_Wv_W,
\end{equation}
of the wind.  

Putting our lower limit on column density $N_{Wabs}>3.4\times 10^{23}\pcmsq$ into eqn.~\ref{eq:mtm} gives
\begin{equation}
P_W>1.4\times 10^{36}\,{\rm dyne}\approx 10\left(\frac{L_{\rm bol}}{c}\right),
\end{equation}
where we have used a black hole mass of $M=2.5\times 10^9\Msun$ and assumed that the wind covers a solid angle of $\Omega=\pi$.  Thus, the momentum flux in the wind exceeds the photon momentum flux by an order of magnitude.   Including the information on the wind velocity also gives us lower bounds on the mass and energy flux (using eqns.~\ref{eq:massflux} and \ref{eq:energyflux}),
\begin{eqnarray}
\dot{M}_W&>&7.7\,\Msunpyr\approx 110\left(\frac{L_{\rm bol}}{c^2}\right),\\
L_K&>&1.7\times 10^{45}\ergps\approx 0.42L_{\rm bol}.
\end{eqnarray}
We can immediately see that the mass flux in the wind exceeds the accretion rate onto the black hole unless the radiative efficiency is $\eta<9\times 10^{-3}$, and that the energy flux is at least 42\% of the bolometric luminosity. These bounds are based on the assumption that the observed wind velocity is the local escape velocity (thereby allowing us to localize the wind in radius).   If,  instead, we made the weaker assumption that the wind is moving at {\it no less than} its local escape velocity, the momentum, mass and energy fluxes further increase.

Given these lower limits on wind fluxes, we can make some comment about acceleration mechanisms.  For radiative-driving, the fact that $P_W>10L_{\rm bol}/c$ would require acceleration of a very Compton-thick wind, $\tau_e>10$, that completely surrounds the radiation source \citep{reynolds:12b}.  While not ruled out by these data, this would require a revision of our basic picture for the central engine structure of AGN.  Magneto-centrifugal acceleration is possible provided that the inward mass flux in the accretion disk exceeds the mass flux in the wind (in order for the disk to generate sufficient torque to generate the wind). To make this compatible with our measured limits on the mass flux requires a low accretion efficiency, $\eta\lesssim 0.01$, as might be expected for these low Eddington ratios due to the formation of an optically-thin advective accretion flow. If Cygnus~A does indeed possess an advective accretion flow, we also expect some contribution to the wind acceleration to come from thermal driving \citep{blandford:99a}.

Of course, Cygnus~A also possesses very powerful relativistic jets.   Using the dynamics of the radio cocoon, \cite{ito:08a} estimate a total jet power in the range $L_j\approx (0.7-4)\times 10^{46}\ergps$; this exceeds the radiative luminosity by a factor of 2--10.  Whatever process drives these relativistic jets, whether it be accretion \citep{blandford:82a} or black hole spin \citep{blandford:77a}, we need only channel $<25\%$ of the power into a wide-angle sub-relativistic outflow in order to explain the observed wind. 

\subsection{Implications for AGN Feedback}

A large body of prior work has identified two distinct modes of AGN feedback \citep{fabian:12a}.  Relativistic radio-jets from AGN in massive ellipticals and the brightest cluster galaxies in cooling core clusters appear to provide ``maintenance mode'' feedback, preventing a cooling catastrophe in their hot atmospheres \citep{peterson:06a}.   On the other hand, powerful sub-relativistic winds from luminous quasars can sweep molecular gas out of a galaxy and quench star formation following a major merger \citep{hopkins:06a,tombesi:15a}.  These are normally considered to be mutually exclusive forms of feedback.   This notion is partially motivated by drawing an analogy between AGN and stellar mass black hole X-ray binaries (BHBs) in which the mutually exclusive occurrence of disk winds and relativistic jets is quite well established \citep{ponti:12a,king:13a}.  

Cygnus~A appears to break this mold.   We clearly see powerful jets interacting with the ICM of the cooling core cluster.   However, we now also see a powerful and wide-angle sub-relativistic wind with sufficient energy and momentum to significantly impact the host galaxy \citep{hopkins:10a}.  We should not be too surprised that some breakdown in the AGN-BHB analogy exists.  Important aspects of the accretion physics do {\it not} remain invariant as we scale the mass up from a stellar-mass to a supermassive object.  AGN accretion disks are significantly more radiation-pressure dominated than their BHB cousins, raising the possibility that their stability properties are different.  Additionally, being cooler, AGN disks have much higher opacities at their photospheres raising the possibility of line-driven winds \citep{laor:14a}.  Line-driving may prime the fast wind that we see, lofting material off the surface of the disk where it can be strongly photo-ionized by the central X-ray source and magneto-centrifugally accelerated.

\acknowledgments
\section*{Acknowledgments}
We thank the anonymous referee for their thorough and constructive comments that improved the quality of the manuscript.   The authors also thank Francesco Tombesi for stimulating conversations throughout the course of this work.  CSR thanks NASA for support under grant NNX14AF86G.   Our analysis makes use of the {\tt XSPEC\_EMCEE} package developed and distributed by Jeremy Sanders.   This work was supported under NASA Contract No. NNG08FD60C, and made use of data from the {\it NuSTAR} mission, a project led by the California Institute of Technology, managed by the Jet Propulsion Laboratory and funded by the National Aeronautics and Space Administration. We thank the NuSTAR Operations, Software and Calibration teams for support with the execution and analysis of these observations. This research has made use of the NuSTAR Data Analysis Software (NuSTARDAS) jointly developed by the ASI Science Data Center (ASDC, Italy) and the California Institute of Technology (USA).

{\it Facilities: }NuSTAR, XMM-Newton, Chandra

\bibliographystyle{jwapjbib}

\begin{thebibliography}{}

\bibitem[\protect\astroncite{{Antonucci}, {Hurt} \&
  {Kinney}}{1994}]{antonucci:94a}
{Antonucci}, R., {Hurt}, T., \& {Kinney}, A.,  1994, \nat, 371, 313

\bibitem[\protect\astroncite{{Arnaud} et~al.}{1984}]{arnaud:84a}
{Arnaud}, K.~A., {Fabian}, A.~C., {Eales}, S.~A., {Jones}, C., \& {Forman}, W.,
   1984, \mnras, 211, 981

\bibitem[\protect\astroncite{{Arnaud} et~al.}{1987}]{arnaud:87a}
{Arnaud}, K.~A., {Johnstone}, R.~M., {Fabian}, A.~C., {Crawford}, C.~S.,
  {Nulsen}, P.~E.~J., {Shafer}, R.~A., \& {Mushotzky}, R.~F.,  1987, \mnras,
  227, 241

\bibitem[\protect\astroncite{{Barthel} \& {Arnaud}}{1996}]{barthel:96a}
{Barthel}, P.~D., \& {Arnaud}, K.~A.,  1996, \mnras, 283, L45

\bibitem[\protect\astroncite{{Basko}}{1978}]{basko:78a}
{Basko}, M.~M.,  1978, \apj, 223, 268

\bibitem[\protect\astroncite{{Belsole} \& {Fabian}}{2007}]{belsole:07a}
{Belsole}, E., \& {Fabian}, A.~C.,  2007,
\newblock in Heating versus Cooling in Galaxies and Clusters of Galaxies, ed.
  H. {B{\"o}hringer}, G.~W. {Pratt}, A. {Finoguenov}, P. {Schuecker},  101

\bibitem[\protect\astroncite{{Blandford} \& {Begelman}}{1999}]{blandford:99a}
{Blandford}, R.~D., \& {Begelman}, M.~C.,  1999, \mnras, 303, L1

\bibitem[\protect\astroncite{{Blandford} \& {Payne}}{1982}]{blandford:82a}
{Blandford}, R.~D., \& {Payne}, D.~G.,  1982, \mnras, 199, 883

\bibitem[\protect\astroncite{{Blandford} \& {Znajek}}{1977}]{blandford:77a}
{Blandford}, R.~D., \& {Znajek}, R.~L.,  1977, \mnras, 179, 433

\bibitem[\protect\astroncite{{Carilli} \& {Barthel}}{1996}]{carilli:96a}
{Carilli}, C.~L., \& {Barthel}, P.~D.,  1996, \aapr, 7, 1

\bibitem[\protect\astroncite{{Dickey} \& {Lockman}}{1990}]{dickey:90a}
{Dickey}, J.~M., \& {Lockman}, F.~J.,  1990, \araa, 28, 215

\bibitem[\protect\astroncite{{Djorgovski} et~al.}{1991}]{djorgovski:91a}
{Djorgovski}, S., {Weir}, N., {Matthews}, K., \& {Graham}, J.~R.,  1991, \apjl,
  372, L67

\bibitem[\protect\astroncite{{Fabian}}{2012}]{fabian:12a}
{Fabian}, A.~C.,  2012, \araa, 50, 455

\bibitem[\protect\astroncite{{Fanaroff} \& {Riley}}{1974}]{fanaroff:74a}
{Fanaroff}, B.~L., \& {Riley}, J.~M.,  1974, \mnras, 167, 31P

\bibitem[\protect\astroncite{{Foreman-Mackey} et~al.}{2013}]{foreman:12a}
{Foreman-Mackey}, D., {Hogg}, D.~W., {Lang}, D., \& {Goodman}, J.,  2013,
  \pasp, 125, 306

\bibitem[\protect\astroncite{{Garc{\'{\i}}a} et~al.}{2014}]{garcia:14a}
{Garc{\'{\i}}a}, J., et~al., 2014, \apj, 782, 76

\bibitem[\protect\astroncite{{Garc{\'{\i}}a} et~al.}{2013}]{garcia:13a}
{Garc{\'{\i}}a}, J., {Dauser}, T., {Reynolds}, C.~S., {Kallman}, T.~R.,
  {McClintock}, J.~E., {Wilms}, J., \& {Eikmann}, W.,  2013, \apj, 768, 146

\bibitem[\protect\astroncite{{George} \& {Fabian}}{1991}]{george:91a}
{George}, I.~M., \& {Fabian}, A.~C.,  1991, \mnras, 249, 352

\bibitem[\protect\astroncite{{Hargrave} \& {Ryle}}{1974}]{hargrave:74a}
{Hargrave}, P.~J., \& {Ryle}, M.,  1974, \mnras, 166, 305

\bibitem[\protect\astroncite{{Harrison} et~al.}{2013}]{harrison:13a}
{Harrison}, F.~A., et~al., 2013, \apj, 770, 103

\bibitem[\protect\astroncite{{Hopkins} \& {Elvis}}{2010}]{hopkins:10a}
{Hopkins}, P.~F., \& {Elvis}, M.,  2010, \mnras, 401, 7

\bibitem[\protect\astroncite{{Hopkins} et~al.}{2006}]{hopkins:06a}
{Hopkins}, P.~F., {Hernquist}, L., {Cox}, T.~J., {Di Matteo}, T., {Robertson},
  B., \& {Springel}, V.,  2006, \apjs, 163, 1

\bibitem[\protect\astroncite{{Ito} et~al.}{2008}]{ito:08a}
{Ito}, H., {Kino}, M., {Kawakatu}, N., {Isobe}, N., \& {Yamada}, S.,  2008,
  \apj, 685, 828

\bibitem[\protect\astroncite{{Kallman} \& {Bautista}}{2001}]{kallman:01a}
{Kallman}, T., \& {Bautista}, M.,  2001, \apjs, 133, 221

\bibitem[\protect\astroncite{{King} et~al.}{2013}]{king:13a}
{King}, A.~L., et~al., 2013, \apj, 762, 103

\bibitem[\protect\astroncite{{Laor} \& {Davis}}{2014}]{laor:14a}
{Laor}, A., \& {Davis}, S.~W.,  2014, \mnras, 438, 3024

\bibitem[\protect\astroncite{{Madsen et al.}}{2015}]{madsen:15a}
{Madsen et al.}, K.,  2015, \apj,  in preparation

\bibitem[\protect\astroncite{{Murphy} \& {Yaqoob}}{2009}]{murphy:09a}
{Murphy}, K.~D., \& {Yaqoob}, T.,  2009, \mnras, 397, 1549

\bibitem[\protect\astroncite{{Ogle} et~al.}{1997}]{ogle:97a}
{Ogle}, P.~M., {Cohen}, M.~H., {Miller}, J.~S., {Tran}, H.~D., {Fosbury},
  R.~A.~E., \& {Goodrich}, R.~W.,  1997, \apjl, 482, L37

\bibitem[\protect\astroncite{{Perley}, {Dreher} \& {Cowan}}{1984}]{perley:84a}
{Perley}, R.~A., {Dreher}, J.~W., \& {Cowan}, J.~J.,  1984, \apjl, 285, L35

\bibitem[\protect\astroncite{{Peterson} \& {Fabian}}{2006}]{peterson:06a}
{Peterson}, J.~R., \& {Fabian}, A.~C.,  2006, \physrep, 427, 1

\bibitem[\protect\astroncite{{Planck Collaboration} et~al.}{2013}]{ade:13a}
{Planck Collaboration}et~al., 2013, ArXiv e-prints

\bibitem[\protect\astroncite{{Ponti} et~al.}{2012}]{ponti:12a}
{Ponti}, G., {Fender}, R.~P., {Begelman}, M.~C., {Dunn}, R.~J.~H., {Neilsen},
  J., \& {Coriat}, M.,  2012, \mnras, 422, L11

\bibitem[\protect\astroncite{{Privon} et~al.}{2012}]{privon:12a}
{Privon}, G.~C., {Baum}, S.~A., {O'Dea}, C.~P., {Gallimore}, J., {Noel-Storr},
  J., {Axon}, D.~J., \& {Robinson}, A.,  2012, \apj, 747, 46

\bibitem[\protect\astroncite{{Protassov} et~al.}{2002}]{protassov:02a}
{Protassov}, R., {van Dyk}, D.~A., {Connors}, A., {Kashyap}, V.~L., \&
  {Siemiginowska}, A.,  2002, \apj, 571, 545

\bibitem[\protect\astroncite{{Reynolds}}{2012}]{reynolds:12b}
{Reynolds}, C.~S.,  2012, \apjl, 759, L15

\bibitem[\protect\astroncite{{Reynolds} et~al.}{2012}]{reynolds:12a}
{Reynolds}, C.~S., {Brenneman}, L.~W., {Lohfink}, A.~M., {Trippe}, M.~L.,
  {Miller}, J.~M., {Fabian}, A.~C., \& {Nowak}, M.~A.,  2012, \apj, 755, 88

\bibitem[\protect\astroncite{{Reynolds} \& {Fabian}}{1996}]{reynolds:96a}
{Reynolds}, C.~S., \& {Fabian}, A.~C.,  1996, \mnras, 278, 479

\bibitem[\protect\astroncite{{Sarazin}, {Finoguenov} \&
  {Wik}}{2013}]{sarazin:12a}
{Sarazin}, C.~L., {Finoguenov}, A., \& {Wik}, D.~R.,  2013, Astronomische
  Nachrichten, 334, 346

\bibitem[\protect\astroncite{{Smith} et~al.}{2002}]{smith:02a}
{Smith}, D.~A., {Wilson}, A.~S., {Arnaud}, K.~A., {Terashima}, Y., \& {Young},
  A.~J.,  2002, \apj, 565, 195

\bibitem[\protect\astroncite{{Smith} et~al.}{2001}]{smith:01a}
{Smith}, R.~K., {Brickhouse}, N.~S., {Liedahl}, D.~A., \& {Raymond}, J.~C.,
  2001, \apjl, 556, L91

\bibitem[\protect\astroncite{{Tadhunter} et~al.}{2003}]{tadhunter:03a}
{Tadhunter}, C., {Marconi}, A., {Axon}, D., {Wills}, K., {Robinson}, T.~G., \&
  {Jackson}, N.,  2003, \mnras, 342, 861

\bibitem[\protect\astroncite{{Tombesi} et~al.}{2015}]{tombesi:15a}
{Tombesi}, F., {Mel{\'e}ndez}, M., {Veilleux}, S., {Reeves}, J.~N.,
  {Gonz{\'a}lez-Alfonso}, E., \& {Reynolds}, C.~S.,  2015, \nat, 519, 436

\bibitem[\protect\astroncite{{Ueno} et~al.}{1994}]{ueno:94a}
{Ueno}, S., {Koyama}, K., {Nishida}, M., {Yamauchi}, S., \& {Ward}, M.~J.,
  1994, \apjl, 431, L1

\bibitem[\protect\astroncite{{Vasudevan} \& {Fabian}}{2009}]{vasudevan:09a}
{Vasudevan}, R.~V., \& {Fabian}, A.~C.,  2009, \mnras, 392, 1124

\bibitem[\protect\astroncite{{Wilms}, {Allen} \& {McCray}}{2000}]{wilms:00a}
{Wilms}, J., {Allen}, A., \& {McCray}, R.,  2000, \apj, 542, 914

\bibitem[\protect\astroncite{{Wilson}, {Smith} \& {Young}}{2006}]{wilson:06a}
{Wilson}, A.~S., {Smith}, D.~A., \& {Young}, A.~J.,  2006, \apjl, 644, L9

\bibitem[\protect\astroncite{{Young} et~al.}{2002}]{young:02a}
{Young}, A.~J., {Wilson}, A.~S., {Terashima}, Y., {Arnaud}, K.~A., \& {Smith},
  D.~A.,  2002, \apj, 564, 176

\end{thebibliography}

\end{document}